\documentclass[fleqn,usenatbib]{mnras}
\usepackage[dvipdfmx]{graphicx}
\usepackage{pdfpages}
\usepackage{multirow}
\usepackage{physics}
\usepackage{float}
\usepackage{booktabs}
\usepackage{amsmath,amssymb,bm}
\usepackage{ascmac}
\usepackage{array}
\usepackage{comment}

\newcommand{\msun}{{\rm M}_\odot}
\newcommand{\mdot}{{\rm M}_\odot\,{\rm yr}^{-1}}

\newcommand{\cc}{{\rm cm^{-3}}}

\newcommand{\vect}[1]{\mbox{\boldmath$#1$}}

\title
[Effects on Mass Accretion Rate]%[**/45 characters]
{Environmental Effects of Star-Forming Cores on Mass Accretion Rate}

\author[Nozaki \& Machida]{Shingo Nozaki$^{1}$ and Masahiro N. Machida$^{1}$ 
\\
$^{1}$
Department of Earth and Planetary Sciences, Faculty of Sciences, Kyushu University, Fukuoka, Fukuoka 819-0395, Japan\\
}
\date{Accepted December 21, 2022}
\pubyear{2022}

\begin{document}
\label{firstpage}
\pagerange{\pageref{firstpage}--\pageref{lastpage}}
\maketitle
\begin{abstract}%[***/200 words for Letters]
We calculate the evolution of cloud cores embedded in different envelopes to investigate environmental effects on the mass accretion rate onto protostars.
As the initial state, we neglect the magnetic field and cloud rotation, and adopt star-forming cores composed of two parts: a centrally condensed core and an outer envelope.
The inner core has a critical Bonnor--Ebert density profile and is enclosed by the outer envelope.  
We prepare 15 star-forming cores with different outer envelope densities and gravitational radii, within which the gas flows into the collapsing core, and calculate their evolution until $\sim 2\times10^5$\,yr after protostar formation.
The mass accretion rate decreases as the core is depleted when the outer envelope density is low. 
In contrast,  the mass accretion rate is temporarily enhanced when the outer envelope density is high
and the resultant protostellar mass exceeds the initial mass of the centrally condensed core.
Some recent observations indicate that the mass of prestellar cores is too small to reproduce the stellar mass distribution. 
Our simulations show that the mass inflow from outside the core contributes greatly to protostellar mass growth when the core is embedded in a high-density envelope, which could explain the recent observations.
\end{abstract}
%%%%%%%%%%%%%%%%%%%%%%%%%%%%%%%%%%%%%%%%
\section{Introduction}
\label{sec:intro}
Observations of nearby star-forming regions have shown that stars form in molecular cloud cores and have various masses at their formation epochs. 
It is essential to clarify the mass distribution of stars, as stars are the most fundamental element of galaxies and hence of the universe. 
\cite{salpeter1955} investigated the mass distribution (or mass function) of main sequence stars in the solar neighborhood, 
and the stellar mass distribution has been investigated in many subsequent observational studies, with the observations showing that the stellar mass distribution is almost the same in any region of the galaxy. 
The universality of the mass distribution (or the initial mass function, IMF) seems to hold in nearby star-forming regions \citep{kroupa2001}. However, the nature of the star-forming environments, such as star formation activity, stellar number density and turbulent intensity, significantly differ in each star-forming region.
Therefore,  the universality of the IMF has been recognized as a great mystery with regard to star formation.

It is difficult to untangle what determines the IMF,
though many researchers have attempted to explain the universality of the IMF. 
\cite{nutter2007} provided a clue to solving the mystery of the IMF.
They pointed out a similarity between the high-mass slope of the IMF and the core mass function (CMF), constructed from the mass distribution of prestellar cores observed in the Orion star-forming regions. 
The similarity implies that the origin of the IMF can be attributed to the CMF. 
\cite{andre2010} clearly showed a similarity between the CMF and the IMF, coinciding within a factor of 0.3 in mass,
implying that a fixed mass fraction ($\sim0.3$) of protostellar cores is converted to a star. 

In a theoretical study, \cite{nakano1995} found that a significant fraction of the mass in the (prestellar) cloud core is ejected by the protostellar outflow, and the fraction depends little on the cloud core mass.
Using numerical simulations, \cite{machida2012} showed that the protostellar outflow expels  $30$--$70$\% of the mass from  the cloud core. 
These theoretical studies indicate that the protostellar outflow can significantly reduce the star formation efficiency to  30(--70)\%. 
Thus, the similarity between the CMF and the IMF confirmed in observations is consistent with the theoretical predictions. 
Therefore, it is considered that the origin of the IMF is associated with the CMF. However, the origin of the CMF has not yet been clarified in either observational or theoretical studies. 

Recently, \cite{takemura2021b,takemura2021a} observed the Orion Nebula Cluster region and identified about 700 dense cores. 
A comparison between the CMF identified in their study and the IMF in the same region showed good agreement, implying that the star formation efficiency is about $100$\%, 
which again deepens the mystery of the origin of the IMF and the CMF.  

In the Orion Nebula Cluster region, the prestellar cores are highly clustered, and the distances between the cores are as small as $\lesssim 0.1$\,pc \citep[see Fig.1 of ][]{takemura2021b,takemura2021a}.  
Thus, each core's gravitational sphere is expected to be severely limited.
Since mass accretion should occur within the gravitational sphere of each core, the protostellar system cannot accumulate gas from an infinitely distant domain in such a highly clustered region. 
On the other hand, dense gas surrounds the cores within so-called filaments.
Thus, the dense ambient matter outside a prestellar core can supply mass to the core \citep{konyves2020,Redaelli2022}.
It is therefore difficult to predict the resultant stellar mass from the observed prestellar cores, and we cannot simply determine the relation between the stellar mass and the core mass. 

It should be helpful to consider isolated prestellar cores, such as Bok globules, as an opposite case to the highly clustered core environment.
\cite{alves2001} showed that the density profile for Bok globule Barnard 68 is well represented by a Bonnor Ebert (B.E.) profile. 
The conditions outside the Bok globules have been clarified in subsequent observations. 
\cite{nielbock2012} has shown that the centrally condensed Bok globule is smoothly connected to the uniform medium outside the core. 
\cite{kandori2005} investigated the (column) density profile for ten prestellar cores or Bok globules. They showed that the density contrast between the cloud center and cloud edge (ambient density) is 10 to 400. 
\cite{roy2014} also confirmed different ambient densities around prestellar cores. 
Thus, the density outside a centrally condensed Bok globule or the density contrast between a cloud center and the ambient medium differs in each core.  
In other words, there is a wide variety of environments for prestellar cores. 
Thus, the observations of isolated starless cores indicate that stars can form in various cores with different ambient densities.  

\cite{motte2001} investigated the density profile for proto- and prestellar cores, focusing on their environment, and showed that the cores observed in the Taurus and Perseus star-forming regions have a B.E. density profile.
They also noted that the environments around the cores significantly differ, in that the cores have considerably different density profiles in the outer part.
%%They concluded that the mass accretion rate for each core depends on the density profile of the core. 
In the Taurus star-forming region, \cite{palmeirim2013} found a core embedded in a very high-density envelope. 
Thus, although the prestellar cores tend to have a B.E.-like density profile around a density peak, they have a variety of environmental conditions (or a variety of surrounding densities).

In principle, it is difficult to identify the actual parent matter that finally accretes onto a protostar within the star formation (or main accretion) timescale ($\lesssim 10^5$\,yr) because we cannot simply determine which parts of the prestellar cloud core and ambient medium contribute to the mass growth of the protostar. 
As described above, matter in a limited space  could fall onto the protostellar system when a prestellar core is embedded in a highly clustered region such as the Orion Nebula Cluster region.
For an isolated core, the protostar could accumulate matter from a nearly uniform medium widely distributed outside the  star-forming core. 
It would be valuable to investigate how to determine the resulting stellar mass in various star-forming environments. 
As \cite{takemura2021b,takemura2021a} discussed, we need to revisit the relation between the CMF and the IMF to clarify star formation at the cloud core scale.   
The star formation efficiency, namely the conversion factor from a prestellar cloud core to a star in mass, is defined as the ratio of the stellar mass to the prestellar core mass. 
Cores are usually identified in radio and near-infrared observations. 
In radio (interferometer) observations, the mass of the prestellar cores  can be underestimated because we cannot precisely detect (or determine) low-density or uniform ambient matter.
Thus, it is not clear whether the cores identified in \cite{takemura2021b,takemura2021a} are actual prestellar cores because only the region around a peak density is identified as the core. 
In other words, they could miss the ambient matter around the cores. 
The possibility of the existence of the ambient matter (or mass reservoir) outside the cores is pointed out in \citet{takemura2021a}. 
If the ambient matter contributes to the protostellar mass growth within a limited time, the mass of the prestellar cores should be underestimated in the observations.

This study focuses on the mass accretion rate and final stellar mass for the cores embedded in different density environments. 
In our past studies \citep{machida2012, machida2013}, we artificially limited the gas inflow within a range. 
However, the properties of prestellar cloud cores, such as the density contrast between the cloud center and edge, the density of the ambient medium and the distance between cores, should differ from core to core.
In addition, we do not know how to relate the core (mass) identified in radio interferometer observations to the stellar mass. 
In this study, we investigate the environmental effects of cores on the mass accretion rate. 
We also discuss the possible star formation efficiency with the simulation results.

The structure of this paper is as follows. 
The numerical settings and methods are described in \S2. The calculation results are presented in \S3.
We discuss the difference in the mass accretion rate for cores with different environments in \S4. 
A summary is presented in \S5.

%%%%%%%%%%%%%%%%%%%%%
\section{Methods}
%%%%%%%%%%%%%%%%%%%%%

%%%%%%%%%%%%%%%%%%%%%
\subsection{Initial clouds and parameter settings}
\label{sec:initial}
%%%%%%%%%%%%%%%%%%%%%
This study investigates the protostellar mass growth in collapsing star-forming cores with different density profiles and gravitational spheres using three-dimensional nested grid simulations.  
We adopt centrally condensed cloud cores as the prestellar cloud cores for the initial conditions.
The density profile for a prestellar core is composed of two parts, the inner core and the outer envelope, based on various star-forming cores observed in isolated and clustered star-forming regions (see \S\ref{sec:intro}).  
The inner part (hereafter the core) is represented by  the critical B.E.  density profile with a central density  $n_c=10^6\,\cc$ and an isothermal temperature $T_{\rm iso} = 10$\,K  \citep[][]{bonnor1956,ebert1955}. 
The outer part (hereafter the outer envelope or the envelope)  is connected to the core and can have a number of density profiles. 
The structure of the prestellar core (core and envelope) is motivated by past near-infrared observations \citep{kandori2005,nielbock2012,roy2014}, as described in \S\ref{sec:intro}.

To construct a B.E. profile, we define the non-dimensional radius $\xi$  as $\xi \equiv r/(c_{s,0}^2 / 4 \pi G \rho_c)^{1/2}$, where $\rho_c = \mu\, n_{\rm H}\, n_{\rm c}$ is  the central mass density, $c_{s,0}$ is the sound speed, $n_{\rm H}$ is the mass of a hydrogen atom and $\mu$ (= 2.33) is the mean molecular weight.
We adopt the following equation to smoothly connect the inner core  (with the critical B.E. profile) to the outer envelope, 
\begin{equation}
\rho  = \rho_{\rm{(\xi_0)}}  \qty[1 - {\rm A}\, \rm{tanh} \qty{B \qty(\xi - \xi_0)}],
\label{tanh}
\end{equation}
where  $\xi_0$ is the critical B.E. radius ($\xi_0=6.45$) and A and B are control parameters that determine the density profile for the outer envelope. 
Values for the parameters A and B in the models used  are listed in Table~\ref{table1}. 
Equation~(\ref{tanh}) indicates that the cores have the same density profile (critical B.E. profile) in the range $\xi<\xi_0$,
while the density profile for the outer envelope ($\xi>\xi_0$) differs according to parameters A and B. 

%%%%%%%%%%%%%%%%%%%%%
% Table 1
%%%%%%%%%%%%%%%%%%%%%
\begin{table*}
 \caption{
Models and parameters. 
Column 1 gives the model name. 
Columns 2 and 3 gives the non-dimensional $r_{\rm g}$  and dimensional $r_{\rm g, pc}$ parameters for the gravitational sphere.
Column 4 gives the cloud mass $M_{\rm all}$. 
Column 5 gives the external (or minimum) number density $n_{\rm ext}$. 
Columns 6 and 7 give the parameters A and B which determine the density profile for the envelope. 
}
 \label{table1}
 \centering
  \begin{tabular}{ccccccc}
   \hline
    Model& $r_{\rm g}$ & $r_{\rm{g,pc}}$\,[pc] & $M_{\rm{all}}(M_{\rm{sun}})$ &$n_{\rm{ext}}$\, [$\cc$] & $A$ & $B$ \\
   \hline
   B1   & 1 &0.0217 &0.65  &$1.1 \cross 10^5$ & -   & -    \\
   B2H1 & 2 &0.0435 &1.25  &$1.5 \cross 10^4$ & 0.9 & 0.3  \\
   B2H2 & 2 &0.0435 &1.48  &$3.4 \cross 10^4$ & 0.7 & 0.35 \\
   B2H3 & 2 &0.0435 &1.70  &$5.3 \cross 10^4$ & 0.5 & 0.5  \\
   B2H4 & 2 &0.0435 &1.98  &$7.4 \cross 10^4$ & 0.3 & 0.9  \\
   B2L1 & 2 &0.0435 &0.72  &$1.1 \cross 10^3$ & 0.99& 2.0  \\
   B2L2 & 2 &0.0435 &0.77  &$1.1 \cross 10^3$ & 0.99& 1.0  \\
   B2L3 & 2 &0.0435 &0.91  &$1.4 \cross 10^3$ & 0.99& 0.5  \\
   B3H1 & 3 &0.0652 &1.79  &$1.1 \cross 10^4$ & 0.9 & 0.3  \\
   B3H2 & 3 &0.0652 &3.00  &$3.2 \cross 10^4$ & 0.7 & 0.35 \\
   B3H3 & 3 &0.0652 &4.21 &$5.3 \cross 10^4$ & 0.5 & 0.5  \\
   B3H4 & 3 &0.0652 &5.50 &$7.4 \cross 10^4$ & 0.3 & 0.9  \\
   B3L1 & 3 &0.0652 &0.77  &$1.1 \cross 10^3$ & 0.99& 2.0  \\
   B3L2 & 3 &0.0652 &0.82  &$1.1 \cross 10^3$ & 0.99& 1.0  \\
   B3L3 & 3 &0.0652 &0.96  &$1.1 \cross 10^3$ & 0.99& 0.5  \\
   \hline
  \end{tabular}
  \label{param}
\end{table*}
%%%%%%%%%%%%%%%%%%%%%

In addition, we use a parameter $r_{\rm g}$, which corresponds to the radius of the gravitational sphere of the star-forming core and limits the accretion region. 
The parameter $r_{\rm g}$ is normalized by the critical B.E. radius $\xi_0$.
As in \cite{machida2013},  we impose gravity (gas self-gravity and point gravity of protostar or sink) only in the region $r<r_{\rm g}$. 
We switch off the gravity and assign a uniform density in the region $r \ge r_{\rm g}$.
We consider neither the gas motion (or rotation) nor the magnetic field in this study, as described below. 
Thus, almost the same conditions are maintained outside the star-forming core ($r>r_{\rm g}$) during the calculation as in our past studies \citep[for detailed settings, see][]{machida2013,machida2020}. 
We adopt $r_{\rm g}=1,$ 2 and 3. 
The non-dimensional parameter $r_{\rm g}$ and the dimensional radius of the gravitational sphere $r_{\rm g,pc}$ are given in Table~\ref{table1}. 
In the models, a star-forming core can have a critical B.E. radius with $r_{\rm g}=1$, twice the critical B.E. radius with $r_{\rm g}=2$ 
or three times the critical B.E. radius with $r_{\rm g}=3$. 
Hereafter, we call the region within $r<r_{\rm g}$ the star-forming core.

%%%%%%
% Fig.1 
%%%%%%
\begin{figure*}
	\begin{tabular}{cc}
  	\begin{minipage}[t]{0.50\hsize}
     	\centering
    	\includegraphics[width=8cm]{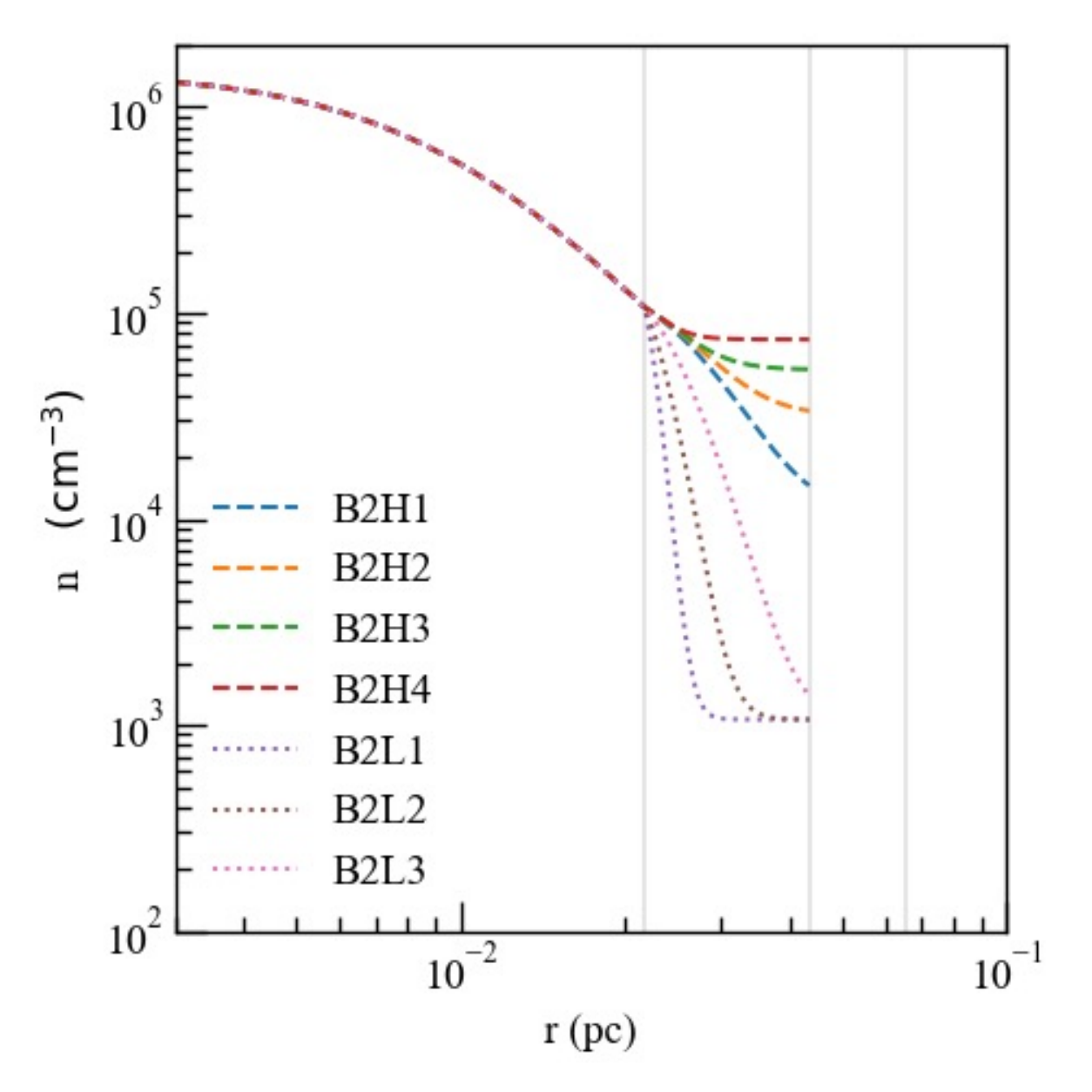}
     \end{minipage} &
     \begin{minipage}[t]{0.5\hsize}
     	\centering
        	\includegraphics[width=8cm]{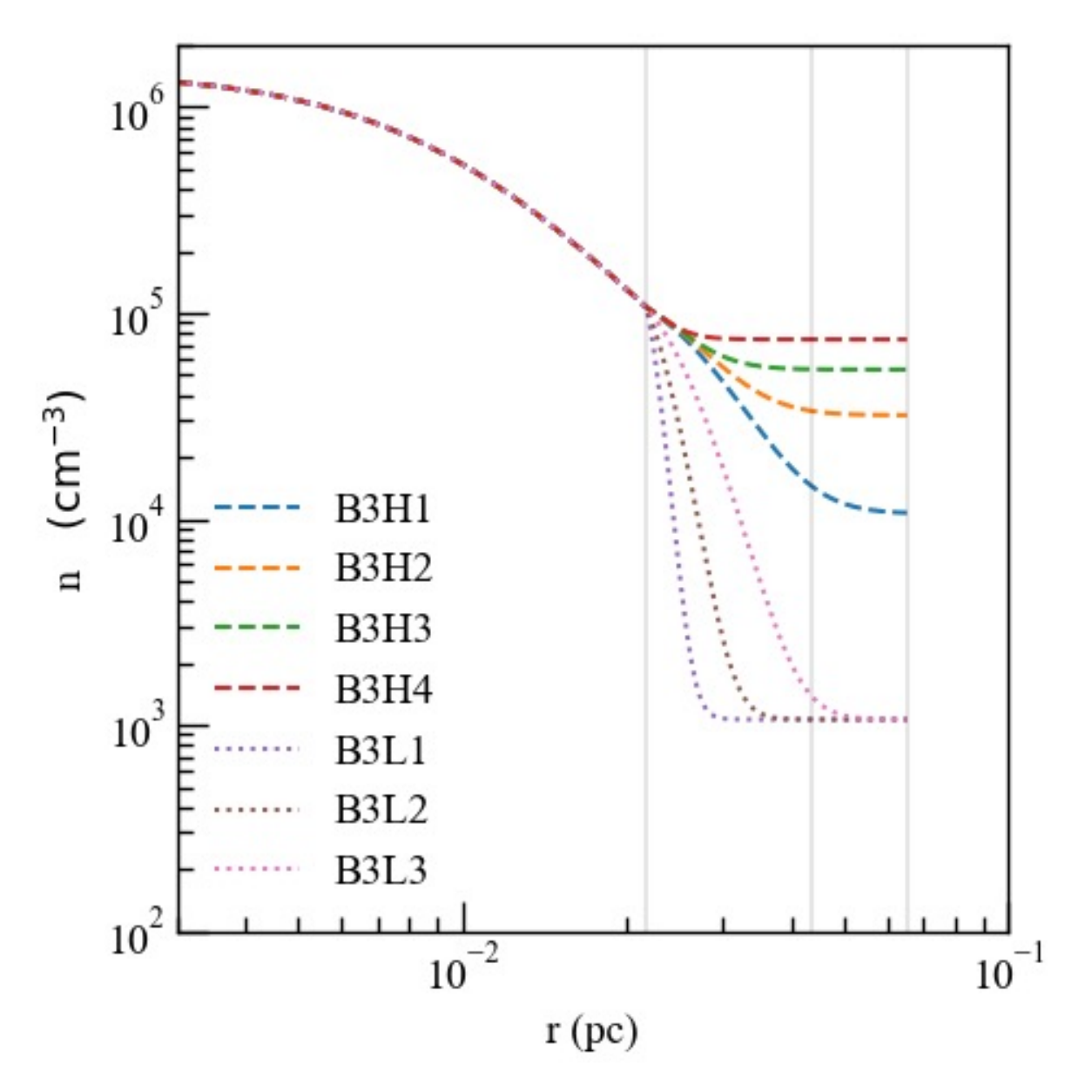}
      \end{minipage}
    \end{tabular}
    \caption{
Density  against radius for models with $r_{\rm g}=2$ (left) and 3 (right). 
The vertical lines in each panel correspond to $r_{\rm g}=1$ (left), 2 (center) and 3 (right), respectively.  }
    \label{density_init}
\end{figure*}
%%%%%%

%%%%%%
% Fig.2
%%%%%%
\begin{figure*}
	\begin{tabular}{cc}
  	\begin{minipage}[t]{0.50\hsize}
     	\centering
    	\includegraphics[width=8cm]{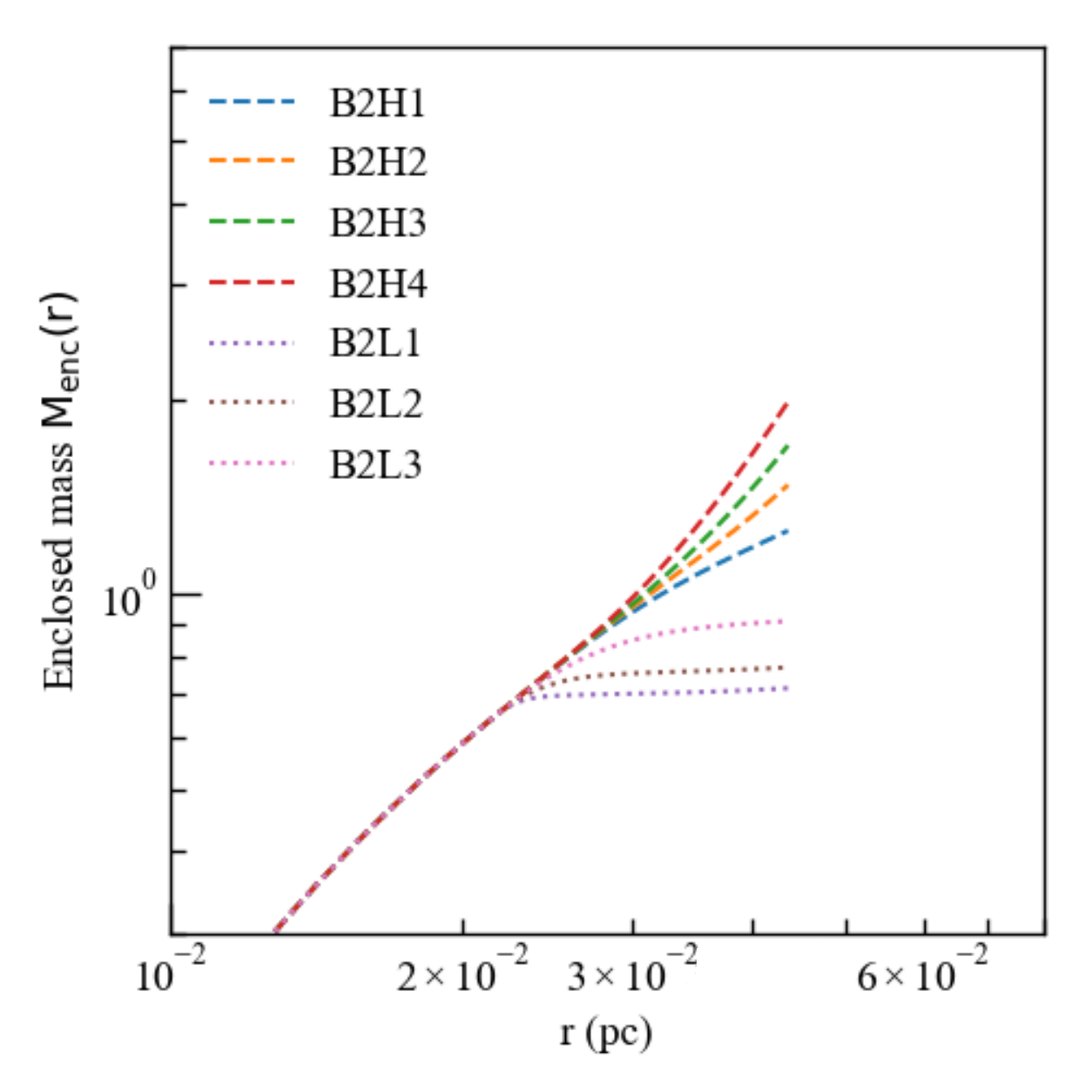}
     \end{minipage} &
     \begin{minipage}[t]{0.5\hsize}
     	\centering
        	\includegraphics[width=8cm]{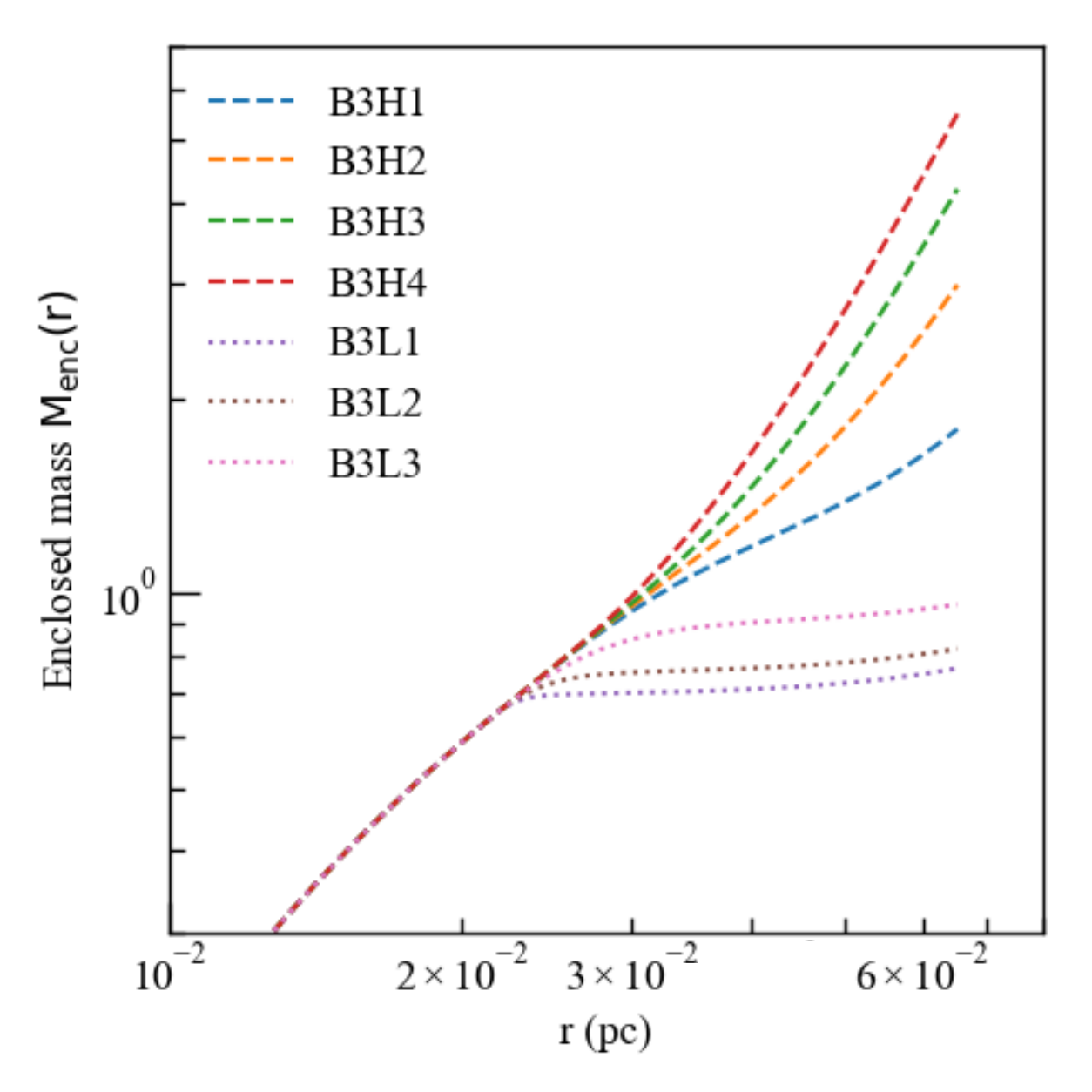}
      \end{minipage}
    \end{tabular}
    \caption{
Enclosed mass against radius for models with $r_{\rm g}=2$ (left) and 3 (right). 
}
    \label{mass_init}
\end{figure*}
%%%%%%

Combining these three parameters, A, B and $r_{\rm g}$, we construct 15 models, as listed in Table~\ref{table1}.  
The density profiles for all models except model B1 are plotted in Figure~\ref{density_init}. 
The left panel of Figure~\ref{density_init} shows  the density profiles for the models with $r_{\rm g}=2$.
As seen in the panel, all the models have the same profile in the range $r\le\xi_0$. 
The density profile in the outer envelope  for model B2H1 is roughly represented by the B.E. density profile with  $\rho\propto r^{-2}$ even in the region $r>\xi_0$.
Models B2H2, B2H3 and B2H4 have a relatively high-density envelope, while models B2L1, B2L2 and B2L3 have a relatively low-density envelope. 
The same trend is seen in the right panel of Figure~\ref{density_init}, in which the models with $r_{\rm g}=3$ are plotted. 
Figure~\ref{density_init} also shows that the ambient density (or the minimum density) differs considerably among the models. 
As described in \S\ref{sec:intro}, we mimic different star formation environments by changing the surrounding (or external) density. 
The external number density $n_{\rm ext}$ in each model is shown  in Table~\ref{table1}.

Figure~\ref{mass_init} shows the enclosed mass $M_{\rm enc}$ for all models against the radius. 
The mass within the gravitational radius for each model is shown in Table~\ref{table1}. 
The mass in the range  $r<\xi_0$ is $M_{\rm all}=0.65\,\msun$, which corresponds to the total mass of model B1. 
We call the mass of model B1 ($M_{\rm core }=0.65\,\msun$) the core mass.  
All the models except B1 have an outer envelope in the range $\xi_0<r<r_{\rm g}$. 
Thus, the models with  $r_{\rm g}=2$ and 3 have a larger total mass than model B1.
The total mass is in the range  0.72 to 1.98\,$\msun$ for the models with $r_{\rm g}=2$ and in the range 0.77 to 5.50\,$\msun$ for the models with $r_{\rm g}=3$. 
For the initial clouds with $r_{\rm g}=2$, the mass ratio of the core  to the outer envelope $M_{\rm env}/M_{\rm core}$ is in the range $0.9$--$2.0$ in  the models with a relatively high-density envelope (models B2H1, B2H2, B2H3, B2H4), in which $M_{\rm env}$ is the mass of the outer envelope in the range $\xi_0 < r < r_{\rm g}$.
The ratio $M_{\rm env}/M_{\rm core}$  is in the range $0.1$--$0.4$ in the models with a relatively low-density envelope (models B2L1, B2L2, B2L3). 
We consider the core in the range  $r<\xi_0$ as the prestellar core identified in observations \citep[e.g.][]{takemura2021b,takemura2021a}.
We could consider the envelope in the range  $r>\xi_0$ as the matter not detected in radio interferometer observations.

%%%%%%%%%%%%%%%%%%%%%
\subsection{Numerical settings}
%%%%%%%%%%%%%%%%%%%%%

We calculate the evolution of the star-forming cores described in \S\ref{sec:initial} using our (M)HD nested grid code with the sink method \citep{machida2004,machida2014,machida2013,machida2020}. 
We prepare nine different grids in order to cover different spatial scales. 
The grid number is described by the index $l$, with the grid size $L(l)$ and cell width $h(l)$ halving with each increment in $l$, i.e.\ $L(l+1)=0.5 L(l) $ and $h(l+1)=0.5 h(l)$. 
Before starting the calculation, we prepare five grid levels (l = 1-5).
After the calculation starts, a new finer grid is automatically generated to ensure that the Truelove condition is met, in which the Jeans length is resolved with at least 32 cells. 
The grid size and cell width for the largest grid ($l=1$) are $L(1)=4.5\times10^5$\,au and $h(1)=7098$\,au,  while those for the finest grid ($l=9$) are  $L(9)=1.78\times10^3$\,au and $h(9)=28$\,au. 

We solve the following basic equations.
\begin{eqnarray} 
& \dfrac{\partial \rho}{\partial t}  + \nabla \cdot (\rho \vect{v}) = 0, & \\
& \rho \dfrac{\partial \vect{v}}{\partial t} 
    + \rho(\vect{v} \cdot \nabla)\vect{v} =
    - \nabla P -      \rho \nabla \phi, & 
\label{eq:eom} \\ 
& \nabla^2 \phi = 4 \pi G \rho, &
\end{eqnarray}
where $\rho$, $\vect{v}$, $P$ and $\phi$ denote the density, velocity, pressure and gravitational potential, respectively. 
We adopt an isothermal polytropic equation of state $P=c_{s,0}^2\, \rho$ to close the equations. 
Note that the magnetohydrodynamics can also be solved with the nested grid code,
though in this study we do not consider the magnetic field so that we could concentrate on the effect of the density of the outer envelope on the mass accretion rate.   
The effect of the inclusion of the magnetic field and feedback effects by outflow are discussed in \S\ref{sec:outflow}. 
We will investigate the outflow feedback in a subsequent study. 

We adopt a sink to accelerate the simulation \citep{machida2010a}. 
The sink cells are generated when the central (or maximum) number density exceeds $10^9\, \cc $.
The threshold density and sink accretion radius are set to $n_{\rm thr}=10^9\,\cc$ and $r_{\rm sink}=78$\,au, respectively. 
With this method, we calculate the evolution of the star-forming cores about $2\times10^5$\,yr after sink creation (or protostar formation). 

%%%%%%%%%%%%%%%%%%%%%
\section{Results}
%%%%%%%%%%%%%%%%%%%%%
\label{sec:results}

%%%%%%%%%%%%%%%%%%%%%
\subsection{Time sequence for typical models}
%%%%%%%%%%%%%%%%%%%%%
%%%%%%
% Fig.3
%%%%%%
\begin{figure*}
	\centering
	\includegraphics[width=12cm]{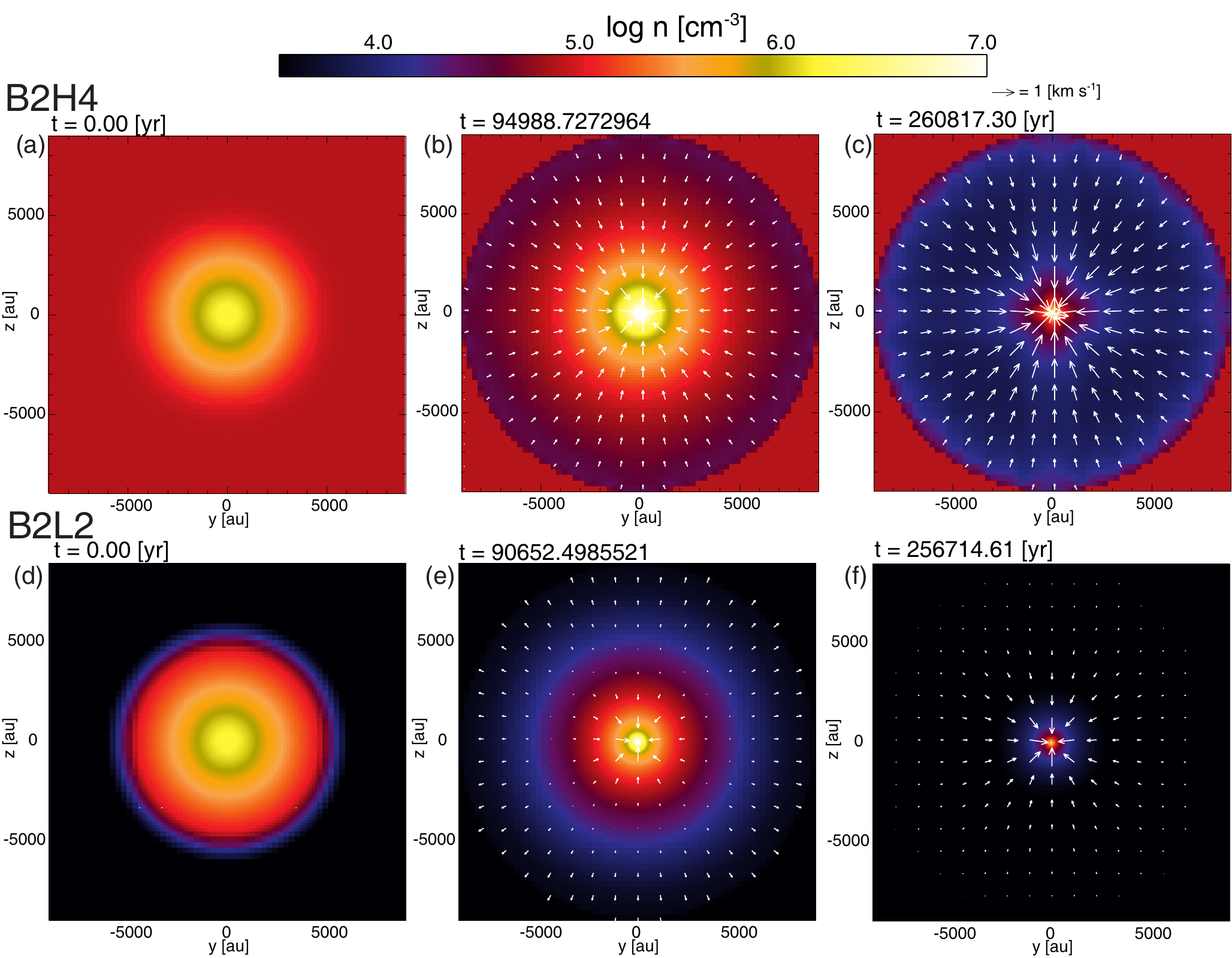}
	\caption{Time sequence for models B2H4 (top) and B2L2 (bottom). 
Density (color) and velocity (arrows) distributions on the $x=0$ plane are plotted in each panel. 
The elapsed time $t$ after protostar formation is shown in the upper left side of each panel.}
	\label{nestedcalc}
\end{figure*}

%%%%%%
% Fig. 
%%%%%%
\begin{figure*}
  \begin{tabular}{cc}
  	\begin{minipage}[t]{0.50\hsize}
     	\centering
     	\includegraphics[width=8cm]{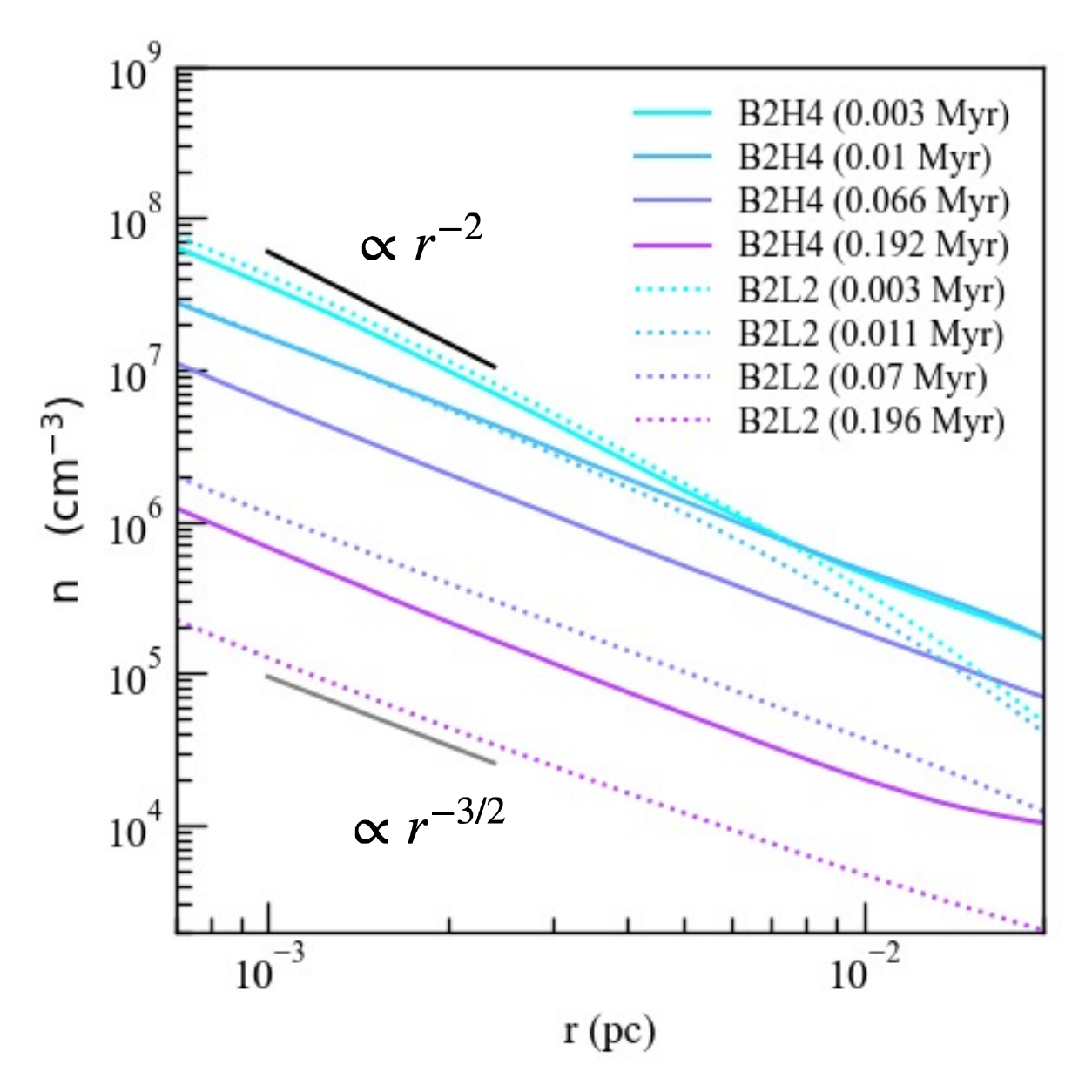}
     \end{minipage} &
     \begin{minipage}[t]{0.50\hsize}
     	\centering
       	\includegraphics[width=8cm]{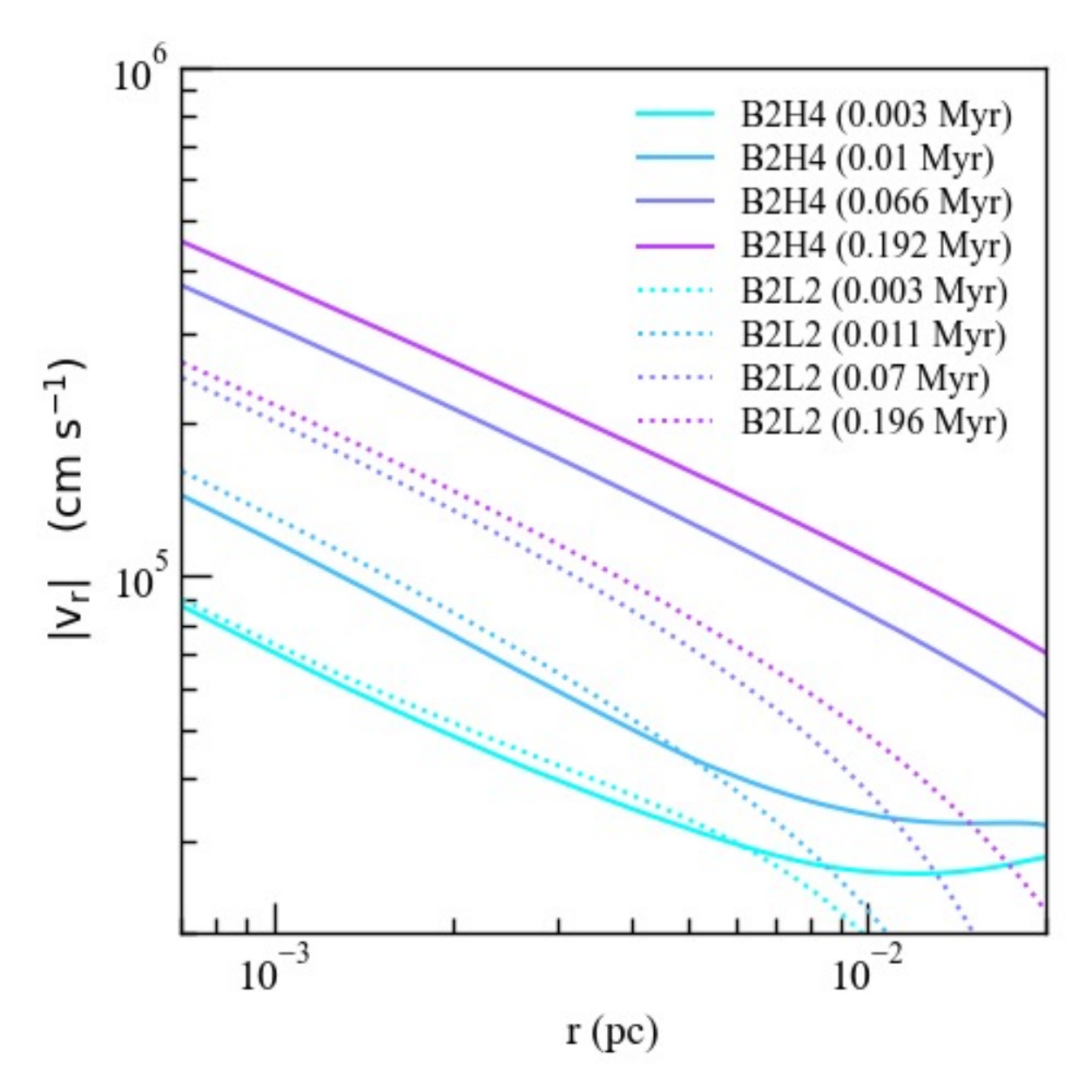}
      \end{minipage}
    \end{tabular}
    \caption{
Density (left) and velocity (right) distributions at four different epochs ( $t \simeq 0.003, 0.01, 0.07, 0.19$\,Myr) against radius for models B2H4 (solid) and B2L2 (dotted). 
}
    \label{rvrrho}
\end{figure*}

We calculate the evolution of the star-forming cores for all models listed in Table~\ref{table1}.
We continue the calculations until the elapsed time after protostar formation exceeds $t \gtrsim 2\times10^5$\,yr.  
As examples, we show the time sequence for two models,  B2H4 (top) and B2L2 (bottom), in Figure~\ref{nestedcalc}.
As shown in Figure~\ref{density_init}, the initial star-forming core for model B2H4 has a relatively high-density envelope, 
while the initial cloud core for model B2L2 has a very low-density envelope. 
The density at the edge of the cloud core is $n_{\rm ext}=7.4 \times 10^4\,\cc$ for model B2H4 and $n_{\rm ext}=1.1\times10^3\,\cc$ for model B2L2  (Table~\ref{table1}). 
We can confirm that the initial star-forming core is embedded in a high-density envelope for model B2H4 in Figure~\ref{nestedcalc}{\it a}.
On the other hand, the initial star-forming core is enclosed by a low-density envelope for model B2L2 (Fig.~\ref{nestedcalc}{\it d}). 
Note that the density color scale is the same among the panels in Figure~\ref{nestedcalc}. 
For these models, we adopted twice the critical B.E. radius as the gravitational radius ($r_{\rm g}=2$), corresponding to $8900$\,au.
Thus, gravitational collapse occurs only in the region $r<8900$\,au. 

Figure~\ref{nestedcalc}{\it b} plots a snapshot at $t\sim10^5$\,yr after protostar formation for model B2H4, and shows that the gas moves toward the center within the gravitational sphere. 
Figure~\ref{nestedcalc}{\it c} shows a snapshot at the end of the calculation for model B2H4, and indicates that the density in the region of $r<r_{\rm g}$ is as low as $\ll10^4\,\cc$ except for the central region.
Thus, the gas within the gravitational sphere has already been depleted by this epoch. 

Figure~\ref{nestedcalc}{\it d} shows that for model B2L2, the density of the outer envelope is very low at the initial state.
Thus, the initial star-forming core for model B2L2 looks like an isolated prestellar core not associated with a dense-gas outer envelope. 
Figure~\ref{nestedcalc}{\it e} plots a snapshot for model B2L2  $\sim 10^5$\, yr after protostar formation, in which the arrows in the region far from the center point outward (i.e., positive radial direction), indicating that the gas in the outer envelope moves away from the center. 
In other words,  part of the envelope gas is outflowing at this epoch.  
As seen in Figure~\ref{density_init}, the density gradient, which is proportional to the pressure gradient, in the range $r>\xi_0$ is large for model B2L2. 
Thus, the envelope gas leaks out from the gravitational sphere. 
However,  the mass ejected from the star-forming core is not large and is within 10\% of the initial mass of the star-forming core,  because the initial mass of the outer envelope is considerably smaller than the total mass of the initial star-forming core, as shown in Figure~\ref{mass_init}.
On the other hand,  all the arrows point to the center (i.e., negative radial direction) at the end of the simulation, indicating that all the gas within the gravitational sphere moves toward the center at this epoch (Fig.~\ref{nestedcalc}{\it f}). 

Figure~\ref{nestedcalc} indicates that the density of the outer envelope becomes very low ($ \ll 10^4\,\cc$) and almost all the gas in the star-forming core has already accreted onto the center by the end of the simulation for both models. 
We can confirm that for both models (B2H4 and B2L2),  the gas in the star-forming core is gradually depleted from the outer envelope in about $10^5$\,yr, and only a small fraction of the mass remains around the center at $\sim2\times10^5$\,yr after protostar formation. 

The density and velocity profiles at four different epochs for models B2H4 and B2L2 are plotted in Figure~\ref{rvrrho}.
For both models, the density profiles in the early main accretion phase ($t \lesssim 0.01$\,Myr) can be described as $\rho \propto r^{-2}$ \citep{larson1969}, while those in the later accretion phase ($t \gtrsim 0.1$\,Myr) obey $\rho\propto r^{-3/2}$  \citep{shu1977}. 
In addition, the density profiles at earlier epochs are almost the same in both models,
but are different at later epochs. 
The density in the infalling envelope is about one order of magnitude higher in model B2H4 than in model B2L2  around the end of the simulation ($t=0.07$--$0.2$\,Myr). 

As shown in the right panel of Figure~\ref{rvrrho}, the velocity profiles at earlier epochs are also almost the same between models B2H4 and B2L2. 
The velocity gradually increases with time (or as the protostellar mass increases). 
We confirmed that the velocities at later epochs correspond well to the freefall velocity, $(2GM_{\rm ps}/r)^{1/2}$, as expected in \citet{shu1977}.
Reflecting the difference in the protostellar mass among the models (for details, see \S\ref{sec:allmodel}), the infall velocity in model B2H4 is higher than in model B2L2.

Both the density and velocity profiles in models B2H4 and B2L2 are almost the same for $t\lesssim0.01$\,Myr. 
A rarefaction wave, which roughly corresponds to the sound speed, should reach the critical B.E. radius $r=\xi_0$ in $0.1$\,Myr (the critical B.E. radius (0.028\,pc) divided by the sound velocity (0.19 km\,s$^{-1}$)),  
and it is considered that the difference in the density and velocity profiles is apparent after the rarefaction wave reaches the critical B.E. radius because the initial density profile for the outer envelope ($r>\xi_0$) differs between the models. 
The difference appears at a somewhat earlier time  (0.028\,Myr) than the sound crossing time ($0.1$\,Myr) due to the infalling motion of the envelope (for details, see \S\ref{sec:previous} and \S\ref{sec:previous2}). 
The difference in the density and velocity profiles in the later main accretion phase should cause a difference in the mass accretion rate and protostellar mass growth, as discussed in the next subsection.

%%%%%%%%%%%%%%%%%%%%%
\subsection{Mass accretion rate and protostellar mass growth for all models}
\label{sec:allmodel}
%%%%%%%%%%%%%%%%%%%%%
%%%%%%
% Fig.4 
%%%%%%
\begin{figure*}
  \begin{tabular}{cc}
  	\begin{minipage}[t]{0.50\hsize}
     	\centering
     	\includegraphics[width=8cm]{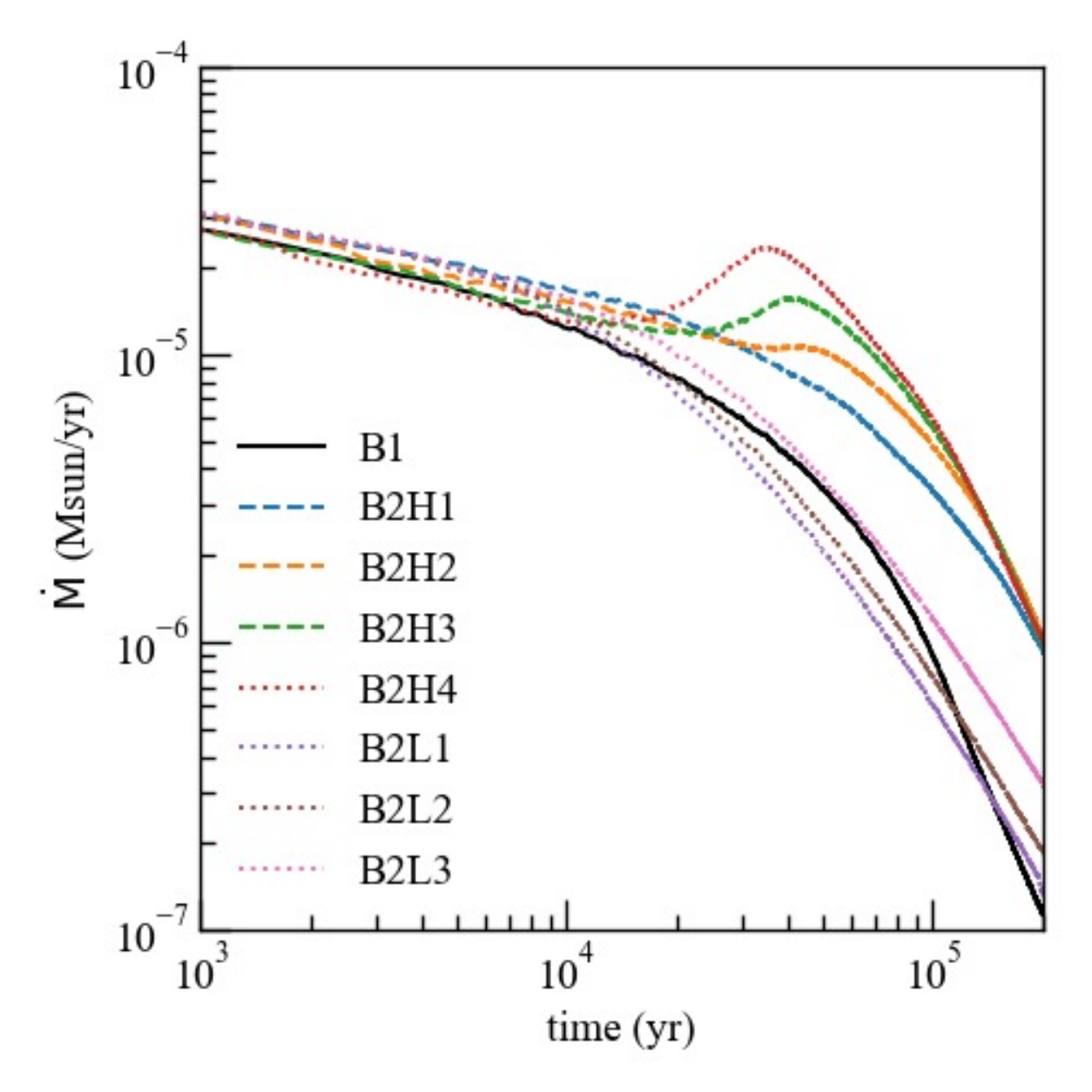}
     \end{minipage} &
     \begin{minipage}[t]{0.50\hsize}
     	\centering
        	\includegraphics[width=8cm]{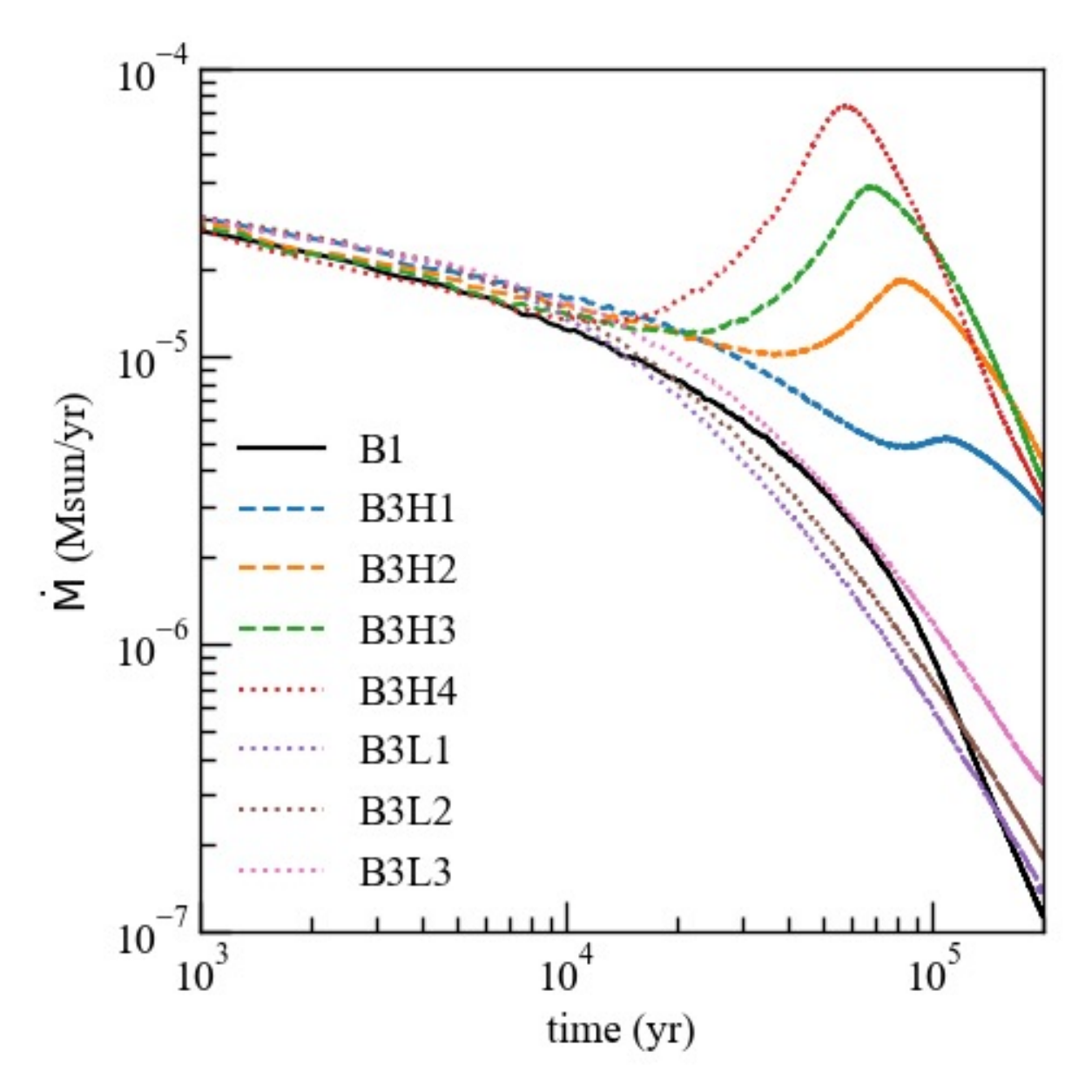}
      \end{minipage}
    \end{tabular}
    \caption{Mass accretion rate for models with $r_{\rm g}=2$ (left) and 3 (right) against the elapsed time after sink formation. 
Model B2 is plotted in both panels for comparison. }
    \label{tmdt}
\end{figure*}
%%%%%%
% Fig. 5
%%%%%%
\begin{figure*}
  \begin{tabular}{cc}
  	\begin{minipage}[t]{0.50\hsize}
     	\centering
     	\includegraphics[width=8cm]{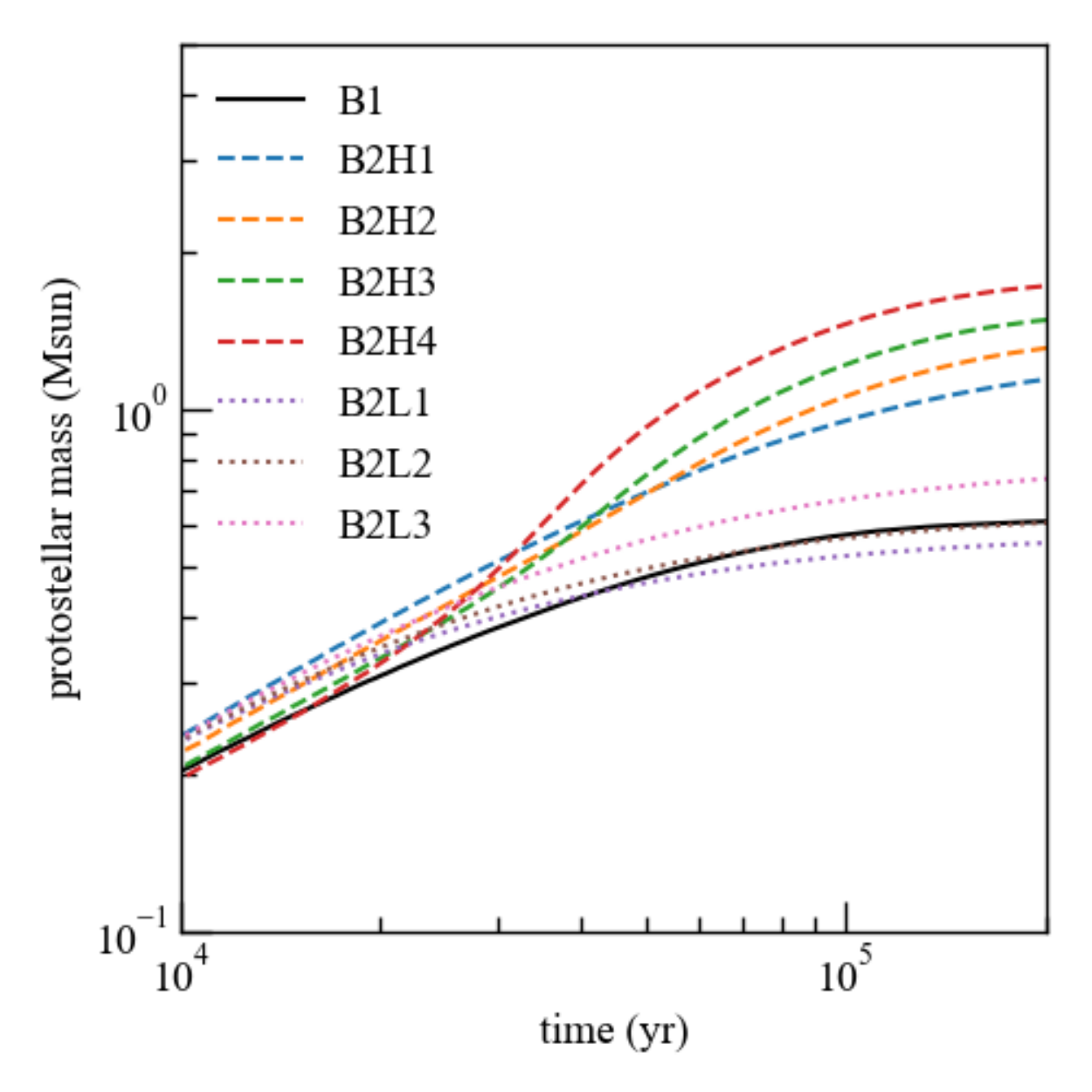}
        	\label{tmdt3Belinear}
     \end{minipage} &
     \begin{minipage}[t]{0.50\hsize}
     	\centering
        	\includegraphics[width=8cm]{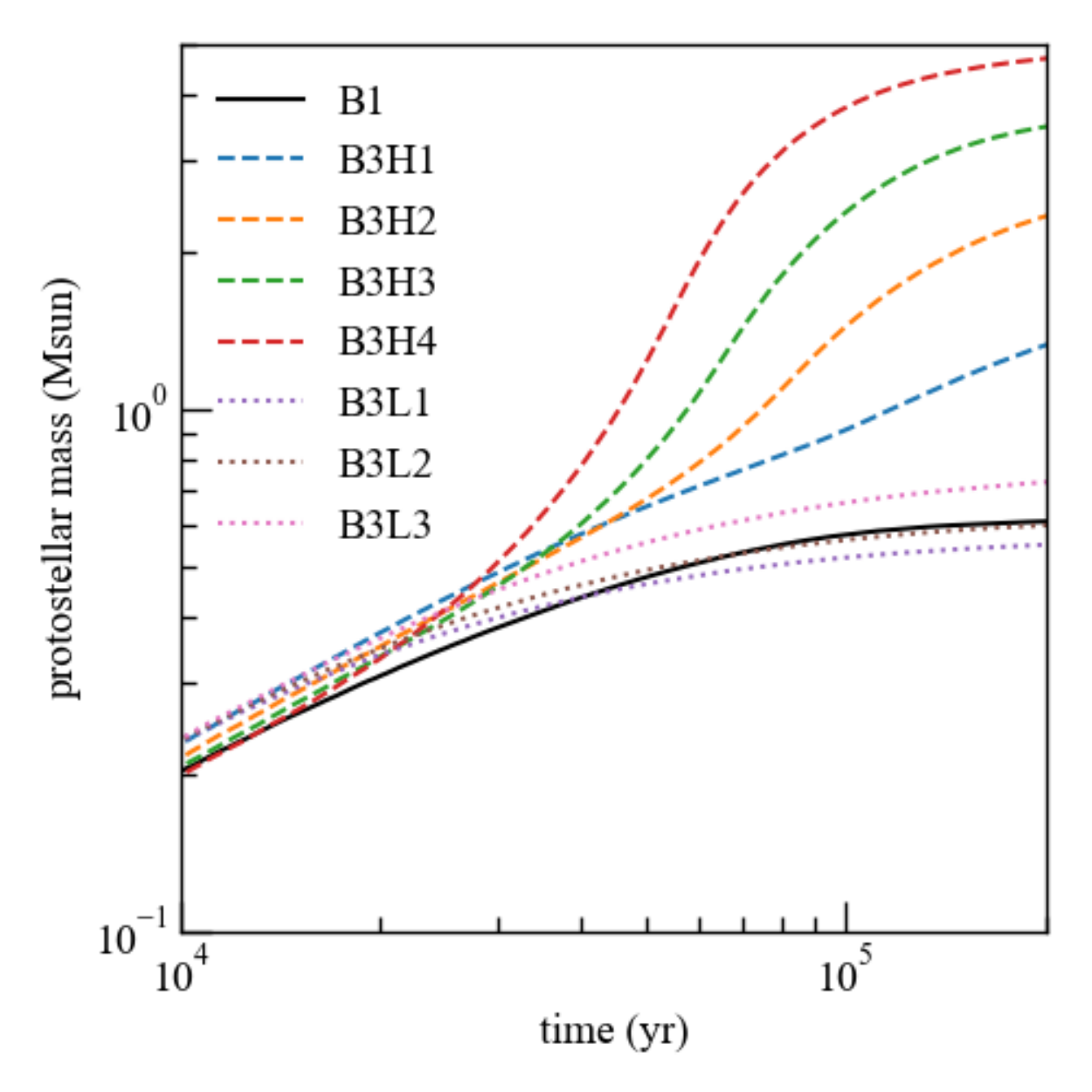}
        	\label{tm3BE}
      \end{minipage}
    \end{tabular}
    \caption{Protostellar mass for models with $r_{\rm g}=2$ (left) and 3 (right) against elapsed time after sink formation. 
Model B2 is plotted in both panels for comparison. }
    \label{tm}
\end{figure*}

Figure~\ref{tmdt} plots the mass accretion rate against the elapsed time after protostar formation for models with $r_{\rm g}=2$ (left) and 3 (right).
In addition, model B1, which has a critical B.E. density profile without an outer envelope, is plotted in both panels for comparison. 
We consider model B1 as the case of an isolated star-forming core. 
Figure~\ref{tmdt} shows that the trend of mass accretion rates among all models are almost the same for $t\lesssim10^4$\,yr, during which the mass accretion rate slightly decreases from $\sim3\times10^{-4}$ to $\sim10^{-5}\,\mdot$. 
The mass accretion process has been investigated in analytical studies \citep[][]{larson2003}.
The mass accretion rate in the self-similar solutions derived from analytical studies can be described as 
\begin{equation}
\frac{dM}{dt} = c_{\rm acc}\, \frac{c_{s.0}^3}{G} \sim  2 \times10^{-6} c_{\rm acc}\, \mdot,
\label{eq:mdot}
\end{equation}
where $c_{\rm acc}$ is the coefficient in the range  $c_{\rm acc}=0.97$--$46.9$ \citep{larson1969, shu1977, hunter1977}. 
Adopting $c_{\rm acc}$ in equation~(\ref{eq:mdot}), the mass accretion rate is in the range  $\dot{M}=(0.2$--$7.4) \times10^{-5}\,\mdot$.
Thus, the mass accretion rates for $t\lesssim10^4$\,yr plotted in Figure~\ref{tmdt} are within the range expected in the self-similar solutions (for more details, see \S\ref{sec:previous}).

Firstly, we focus on the isolated model (model B1). 
The mass accretion rate for model B1 continues to decrease by the end of the simulation and becomes $\dot{M} \lesssim 10^{-7}\, \mdot$ at $t\sim2\times10^5$\,yr (Fig.~\ref{tmdt}). 
The decreasing trend of the mass accretion rate is discussed in \S\ref{sec:previous}. 
The protostellar masses (or the sink masses) for the same models as in Figure~\ref{tmdt} are plotted in Figure~\ref{tm}. 
For model B1, the protostellar mass increases to reach $M\sim 0.5\,\msun$ at the end of the simulation. 
For model B1, since the mass increase rate for $\gtrsim 10^5$\,yr is very small (Fig.~\ref{tm}), a further increase in mass is not expected in subsequent evolutionary stages (see also Fig.~\ref{tmdt}).

Next, we describe the models with a high-density envelope (models B2H1, B2H2, B2H3, B2H4) in the left panel of Figure~\ref{tmdt} (i.e., the models with $r_{\rm g}=2$). 
For these models, the mass accretion rates are temporarily enhanced around  $t\sim (2$--$4) \times10^4$\,yr before they decrease again. 
The mass accretion rates for these models are about one order of magnitude higher than that for  model B1 for $t\gtrsim 2\times 10^4$\,yr. 
Reflecting an increase in the mass accretion rate, the protostellar mass is larger in these models than in model B1 (Fig.~\ref{tm} left panel). 
At the end of the calculation,  the protostellar masses in these models exceed $M\sim1\msun$ and are much larger than that in model B1 ($M\sim0.5\msun$). 

Finally, we focus on the models with a low-density envelope (models B2L1, B2L2, B2L3).
The time evolutions of the mass accretion rate and protostellar mass for these models show almost the same trend as for model B1. 
The left panel of Figure~\ref{tmdt} indicates that the mass accretion rate for models B2L1, B2L2 and B2L3 roughly traces that for model B1. 
The mass accretion rate is slightly larger in these models than in  model B1 at the end of the simulation ($t=2\times10^5$\,yr), because the low-density envelope is considered to affect the mass accretion rate slightly. 
The mass evolution for models B2L1, B2L2 and B2L3 is almost the same as for model B1 (Fig.~\ref{tm} left panel). 
The difference in the protostellar masses among models B1, B2L1, B2L2 and B3L2 is within a factor of 1.5. 

The same trend can be seen in the models with $r_{\rm g}=3$ (right panels of Figs.~\ref{tmdt} and \ref{tm}). 
The enhancement of the mass accretion rate is more significant for the models with a high-density envelope (models B3H1, B3H2, B3H3, B3H4) than for model B1 especially in the later accretion stage ($t \gtrsim 2\times10^4$\,yr). 
As seen in Figure~\ref{tmdt}, among the models with a high-density envelope, the enhancement of the mass accretion rate is considerably larger in  the models with $r_{\rm g}=3$ (models B3H1, B3H2, B3H3, B3H4) than in  the models with $r_{\rm g}=2$ (models B2H1, B2H2, B2H3, B2H4). Among these models, the protostellar mass for models B3H1, B3H2, B3H3 and B3H4 is much larger than that for models B2H1, B2H2, B2H3 and B2H4, as shown in Figure~\ref{tm}. 
The enhancement of the mass accretion rate and protostellar mass can be confirmed in \citet{myers2011}, in which he analytically estimated the resultant (proto)stellar mass in the core with and without a high-density envelope.  
On the other hand, we cannot see a noticeable difference in the mass accretion rate (Fig.~\ref{tmdt})  and protostellar mass (Fig.~\ref{tm}) among the models with a low-density envelope (models B2L1, B2L2, B2L3, B3L1, B3L2, B3L3).  

%%%%%%%%%%%%%%%%%%%%%
\subsection{Apparent star formation efficiency}
%%%%%%%%%%%%%%%%%%%%%
%%%%%%
% Fig. 6
%%%%%%
\begin{figure*}
  \begin{tabular}{cc}
  	\begin{minipage}[t]{0.50\hsize}
     	\centering
     	\includegraphics[width=8cm]{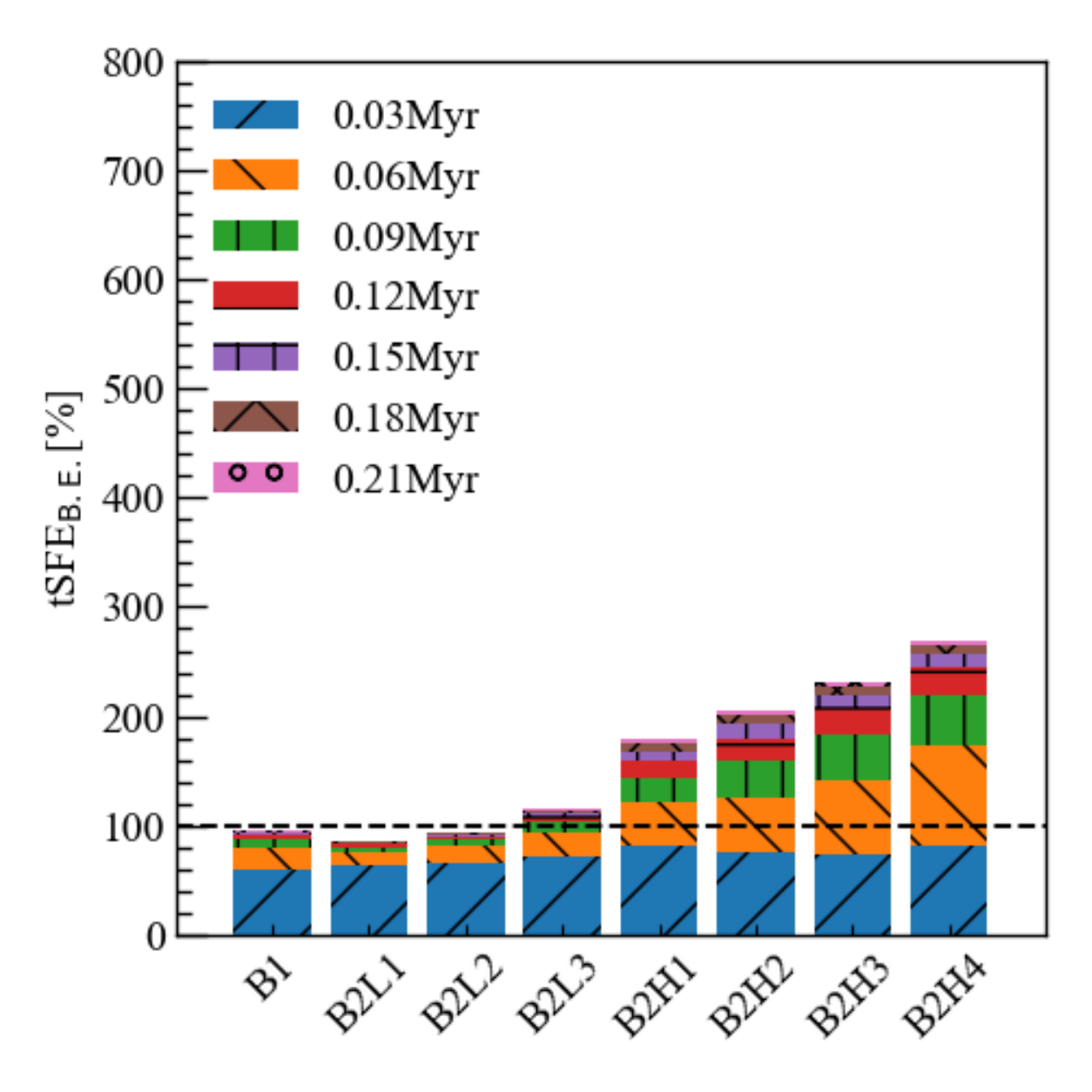}
     \end{minipage} &
     \begin{minipage}[t]{0.50\hsize}
		\centering
		\includegraphics[width=8cm]{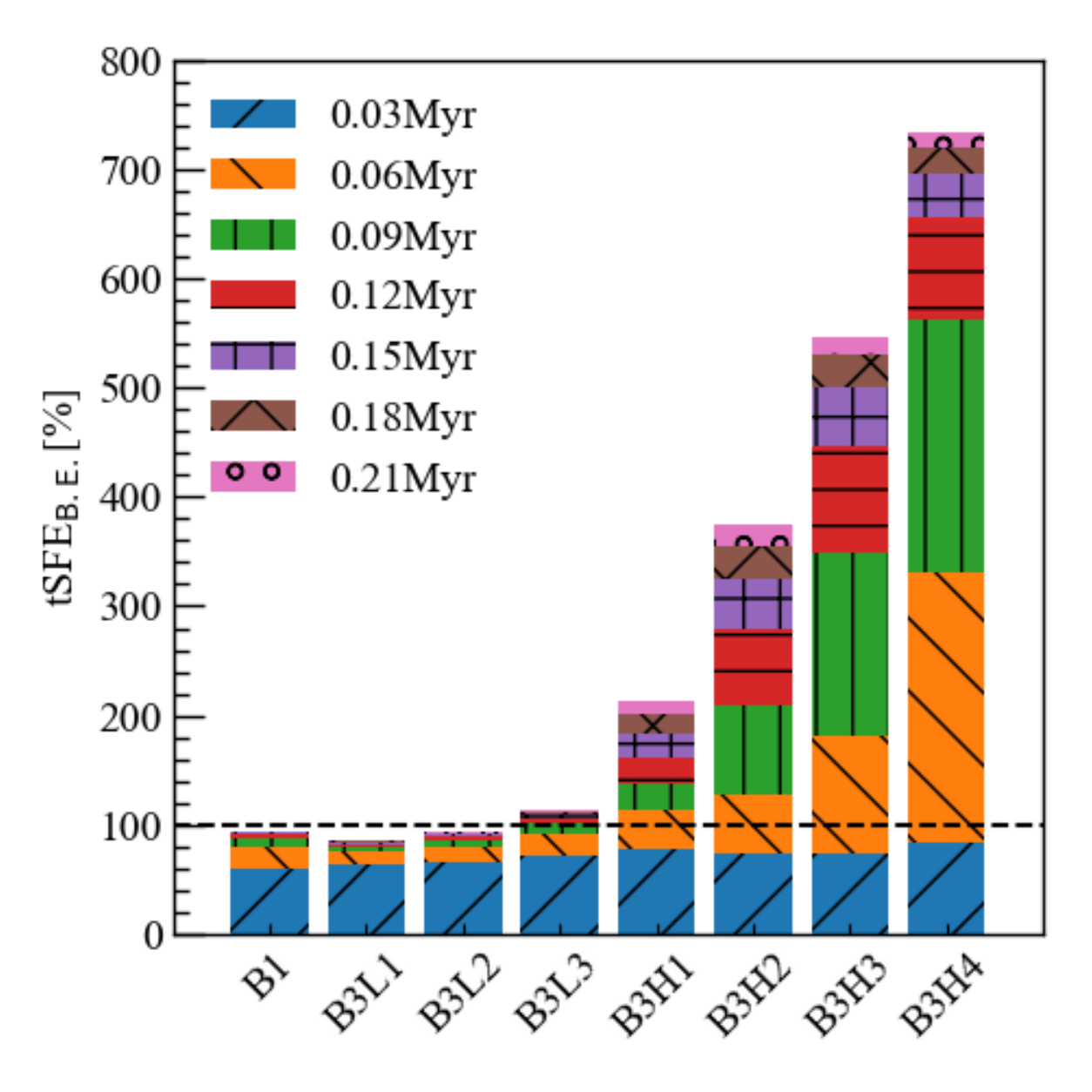}
      \end{minipage}
    \end{tabular}
    \caption{
 Time variable  ${\rm{ tSFE_{B.E.}}}$ (apparent star formation efficiency) every 0.03\,Myr for models with $r_{\rm g}=2$ (left) and 3 (right). 
Model B1 is plotted in both panels for comparison.
}
\label{sfe}
\end{figure*}

The star formation efficiency is usually defined as the (final) protostellar mass divided by the initial cloud core mass. 
In this study, to examine the time evolution of the ratio of the stellar mass to the initial cloud core mass, we define the time variable (apparent) star formation efficiency ($\rm t SFE_{{B. E.}}$) as 
\begin{equation}
	{\rm tSFE_{{B. E. }} [\%]} = \frac{M_{{\rm star}}(t)}{M_{{\rm core}}(t=0)} \times 100,
 \label{sfeeq}
\end{equation}
where $M_{\rm star} (t)$ is the protostellar mass (or sink mass)  at the epoch of $t$, $M_{\rm core}(t=0)$  is the initial core mass for model B1, and we adopt $M_{\rm core}(t=0)=0.65\msun$.
Note that we assume that the prestellar cloud core can be observed as only the high-density region within $r<\xi_0$, as described in \S\ref{sec:cores}. 
Thus, we normalized the protostellar mass by the initial mass of the prestellar cloud core with the critical B.E. radius (i.e., the core) in equation~(\ref{sfeeq}). 
The protostellar mass exceeds the mass of the core  with the critical B.E. density profile ($M_{\rm core}(t=0) =0.65\msun$), because the initial core mass for the models with $r_g=2$ and 3 is higher than $0.65\msun$ (Table~\ref{table1}).  
As a result, the $\rm {t} SFE_{{B. E.}}$ can exceed 100\,\% for the models with $r_g=2$ and 3 because the protostellar mass can exceed the mass of the core.
The definition of $\rm {t} SFE_{{B. E.}}$ used in this paper differs from the star formation efficiency usually  used.  
We call $\rm {t} SFE_{{B. E.}}$ the apparent star formation efficiency.

Figure~\ref{sfe} shows the $\rm {t} SFE_{{B. E.}}$ for the models with $r_{\rm g}=2$ (left panel) and 3 (right panel), with model B1 plotted in both panels for comparison.
In the figure, we stacked the $\rm {t} SFE_{{B. E.}}$ for each 0.3 Myr period to see the time variation.
Since we did not include feedback effects, such as protostellar outflows,  all the mass within the gravitational radius can finally accrete. 
For model B1, which has $r_{\rm g}=1$ (or the critical B.E profile), the $\rm {t} SFE_{{B. E.}}$ ultimately reaches almost 100\%, which is natural because the protostellar mass is normalized by the mass of the core (or the critical B.E sphere). 
Figure~\ref{sfe} also indicates that the mass accretion for model B1 is almost completed within 0.06\,Myr.   

The $\rm {t} SFE_{{B. E.}}$ exceeds 100\% for the models with a high-density envelope because the mass within the gravitational radius  is larger than the mass of the critical B.E. sphere (see Table~\ref{table1}), as described above.  
Figure~\ref{sfe} (left panel) shows that the $\rm {t} SFE_{{B. E.}}$ at $t=0.21$\,Myr is in the range $150$--$250$\% for the models with a high-density envelope (models B2H1, B2H2, B2H3, B2H4) when the gravitational sphere is twice the critical B.E. radius ($r_{\rm g}=2$). 
When the gravitational sphere is three times the critical B.E. radius ($r_{\rm g}=3$), $\rm {t} SFE_{{B. E.}}$ for the models with a high-density envelope (models B3H1, B3H2, B3H3, B3H4) is in the range $200$--$700$\%.

Figure~\ref{sfe} also shows that the (main) accretion phase for the models with a high-density envelope is longer than that for the models with a low-density envelope. 
The main accretion phase for the models with a low-density envelope ends roughly within $\sim0.03$\,Myr,
while a significant fraction of the mass of the star-forming core accretes onto the protostar during $0.06$--$0.12$\,Myr for the models with a high-density envelope.  
Thus, the main accretion phase for the models with a high-density envelope is $2$--$4$ times longer than that for the models with a low-density envelope. Although the duration of the main accretion phase differs among the models plotted in Figure~\ref{sfe}, the main accretion phase ends within 0.15\,Myr even in the models with the highest density envelope (models B2H4 and B3H4). 

%%%%%%%%%%%%%%%%%%%%%%%%%%%%%%%%%%
\section{Discussion}
%%%%%%%%%%%%%%%%%%%%%%%%%%%%%%%%%%

%%%%%%%%%%%%%%%%%%%%%%%%%%%%%%%%%%
\subsection{Comparison of mass accretion rate with previous studies} 
\label{sec:previous}
%%%%%%%%%%%%%%%%%%%%%%%%%%%%%%%%%%

We presented the mass accretion rate and protostellar mass growth when the prestellar cloud cores are embedded in different environments in \S\ref{sec:results}. 
%%The evolution of the mass accretion rate when cloud collapse occurs due to self-gravity has been investigated in many studies.  
The gravitational collapse of an isolated isothermal sphere has been investigated in analytical studies. 
\cite{larson1969} and \cite{penston1969} clarified  the gravitational collapse process for an initially uniform sphere and showed that a self-similar solution can represent the evolution of collapsing cloud cores until protostar formation.

\cite{shu1977} investigated the gravitational contraction process for a singular isothermal sphere (SIS), which has a density profile of  $ \rho(r) = A\, c_s^2  /(4 \pi G r^2)$, where $A$ is a constant.
They showed that the matter in the region that the rarefaction wave passes through could fall on the protostar with the freefall velocity. 
They also pointed out that the mass accretion rate could be described as  $\dot{M}=0.975\, c_s^3/G$, independent of time, which has been confirmed in subsequent studies \citep{hunter1977,whitworth1985}. 
We should note that the relatively low accretion rate with a coefficient of 0.975 predicted by the SIS is attributed to the initial condition without infalling gas motion before protostar formation. 
The coefficient is enhanced when the gas inflow before protostar formation is considered \citep{larson2003}.

With the collapse process of an isothermal sphere clarified through analytical studies, researchers have investigated cloud collapse in numerical simulations.
Using a one-dimensional hydrodynamics code, \cite{foster1993} calculated the evolution of gravitationally collapsing cores with different cloud radii and found that the mass accretion rate decreases shortly after protostar formation when the initial cloud radius is small.
%%They also showed that the mass accretion rate approaches a constant value derived from \cite{shu1977}, while decreasing slightly when the initial cloud core has a large radius.

Using a two-dimensional magnetohydrodynamic code, \cite{tomisaka1996} investigated the evolution of clouds with a parameter of density contrast between the center and the outer edge of the core, and showed that the mass accretion rate converges to a constant value after it rapidly decreases just after protostar formation \citep[see alse][]{ogino1999}. 
The mass accretion rate derived in \cite{tomisaka1996} is higher than that in \cite{shu1977} because the infall motion before protostar formation was considered. 
The difference in the mass accretion rate between the analytical \citep{shu1977} and numerical studies is attributed to the condition of the prestellar core, as described above. 
The infall motion before protostar formation is ignored in \cite{shu1977}, while it is naturally included in numerical studies, 
so the mass accretion rate is larger in the numerical studies than in Shu's solutions. 

\cite{vorobyov2005} investigated the evolution of giant molecular cloud cores using one-dimensional hydrodynamic simulations, focusing on the decrease of the mass accretion rate.  
They showed that the main accretion phase can be divided into three stages depending on the temporal trend of the mass accretion rate. 
After the mass accretion rate rapidly decreases (the first stage), it asymptotically approaches a constant value representing Shu's self-similar solution (the second stage). 
Then,  the mass accretion rate decreases in the final stage.
Two effects cause the decrease of the accretion rate. 
One is the fact that a different inflow treatment used in between  Shu's self-similar solution and the numerical calculations. 
The other is the depletion of the cloud core mass. 

The self-similar solutions roughly represent the mass accretion rate. 
However, there is a limitation in reproducing a realistic mass accretion rate with self-similar solutions. 
The mass accretion rate derived in the self-similar solution can be applied only in a limited evolution time because the effect of a limited mass reservoir cannot be included. 
Using hydrodynamics simulations, a more realistic mass accretion rate has been investigated in previous studies.
The simulation studies have clarified the following: (1) the mass accretion rate is not constant but is time variant, (2) the initial and boundary conditions for the prestellar cloud can affect the mass accretion rate and (3) the mass accretion rate depends on the parameters of the initial cloud core.
The temporal trends of the mass accretion rate for the models with a low-density envelope in this study are in good agreement with those in previous studies \citep[e.g.][]{foster1993,vorobyov2005}, in which the mass accretion gradually decreases with time after it maintains almost the same value given by the self-similar solutions. 
However, the history of the mass accretion rate for the models with a high-density envelope are not represented by previous studies because we considered different environments around star-forming cores.

%%%%%%%%%%%%%%%%%%%%%%%%%%%%%%%%%%
\subsection{Enhancement of mass accretion rate for models with a high-density envelope}
\label{sec:previous2}
%%%%%%%%%%%%%%%%%%%%%%%%%%%%%%%%%%

In some models, the mass accretion rate increases after it initially decreases, as shown in Figure~\ref{tmdt}. 
The mass accretion rate then decreases again in later stages.
The increase of the mass accretion rate also increases the protostellar mass.
We first consider the enhancement of mass accretion in terms of a self-similar solution. 
Assuming a SIS with a density profile of $ \rho \propto c_{s,0}^2\,  r^{-2}$ \citep{shu1977}, the mass $M$ and mass accretion rate $\dot{M}$ are proportional to $M \propto c_{s,0}^2\, r$ and  $ \dot{M} \propto c_{s,0}^2 (dr/dt) \propto c_{s,0}^3$, respectively, in which $dr/dt \sim c_{\rm s,0}$ is adopted for simplisity. 
Since the B.E. sphere has a similar density profile of $\rho \propto r^{-2}$ in the outer region, the mass accretion rate can be roughly represented by Shu's solution (see also \S\ref{sec:previous}).
However, the mass accretion rate should be changed when a different density profile outside the core is considered.
For example, the mass accretion rate is expected to be enhanced with a density profile shallower than the SIS, such as $\rho \propto r^{-2+\alpha}$ ($0<\alpha<2$) in the outer region ($r$ or $\xi \gg 1$), because the envelope mass in a shallower density profile is larger than in a steeper profile ($\alpha<0$) and the velocity of the rarefaction weve (or sound speed) is not changed under isothermal condition. 
%%we can naively expect of mass rethe mass accretion rate as $\dot{M} \propto c_{s,0}^3\, r^\alpha$.
%%Thus, it is natural that the mass accretion rate is enhanced with a shallower density profile.

In the simulations, the epoch at which the enhancement of mass accretion rate occurs corresponds to the second stage defined in  \cite{vorobyov2005}. 
As described above, our simulations showed that the mass accretion rate is enhanced in the intermediate or later main accretion stage when the high-density envelope encloses the prestellar core. 
This trend differs from past studies, and the enhancement of the mass accretion rate should be attributed to the high-density environment we imposed.

The time evolution of the mass accretion rate in isolated collapsing cores has also recently been studied using two or three-dimensional numerical simulations  \citep[e.g.][]{vorobyov2005b,machida2013} to clarify the multi-dimensional nature, such as disk accretion, mass ejection and episodic accretion. 
In such simulations, the mass accretion is not steady due to disk  gravitational instability, and the accretion rate shows a rapid increase in a very short time \citep{vorobyov2006,machida2010a,tomida2017}. 
However, as a whole, the mass accretion rate gradually decreases in a later stage without exhibiting the long-term enhancement of the mass accretion rate shown in the models with a high-density envelope (Fig.\ref{tmdt}).
Although the mass accretion rate suddenly increases due to the disk gravitational instability in some past studies, there is no significant difference in the time evolution of the mass accretion rate between spherical symmetric (or our study) and full three-dimensional simulations when the density of the outer envelope is low. 
In our study, the models with a low-density envelope show a time-decreasing mass accretion rate (Fig.~\ref{tmdt}).  
On the other hand, the models with a high-density envelope show an enhancement of the mass accretion rate in the later stage.
The rate of increase of the mass accretion rate is significant when the gravitational radius is large (i.e., $r_{\rm g}=3$). 
%%Thus, the increase of the mass accretion rate is suggestive of another physical origin. 

%%%%%%%%%%%%%%%%%%%%%%%%%%%%%%%%%%
\subsection{Bondi accretion}
%%%%%%%%%%%%%%%%%%%%%%%%%%%%%%%%%%
Although we can understand the enhancement of the mass accretion rate with self-similar solutions (\S\ref{sec:previous2}), it may be valuable to evaluate the mass accretion rate from the outer envelope in terms of Bondi accretion. 
Steady accretion toward a point mass without the self-gravity of ambient matter is assumed in considering Bondi accretion \citep[][]{bondi1944,bondi1952}. 
In our simulations, the protostellar mass dominates the mass of surrounding gas in a later stage, during which  a nearly steady state is realized (for details, see \S\ref{sec:applicability}). 
Thus, Bondi accretion may explain the mass accretion rate from the outer envelope surrounding the core (or the critical B.E. sphere).
The Bondi accretion rate is represented by 
\begin{equation}
\frac{dM_{\rm b}}{dt} = \frac{4.48 \pi G^2 M_{\rm ps}^2\, \rho_{\rm{\infty}}}{c_{s,0}^3},
\label{bondimdt1}
\end{equation}
where $\rho_{\rm{\infty}}$ is the surrounding (uniform) density.

Since the Bondi accretion rate is proportional to the square of the point mass, it can dominate the mass accretion rate expected in an isolated self-gravitating cloud core, as the protostellar mass increases when the outer envelope is considered.  
To derive the Bondi accretion rate (eq.~[\ref{bondimdt1}]), we introduced the protostellar mass $M_{\rm ps}$, which corresponds to the sink mass.
The isothermal sound speed described in \S\ref{sec:initial} is also used in equation~(\ref{bondimdt1}). 
There is no straightforward way to determine the uniform ambient density $\rho_{\rm \infty}$ in our simulations, because the density far from the protostar slowly changes over time, as seen in Figure~\ref{nestedcalc}. 
Thus, as for $\rho_\infty$, we adopted two different values, (1) the minimum density at each epoch within the gravitational sphere $r<r_{\rm g}$  as $\rho_{\rm \infty}=\rho_{\rm min,t}$, and  (2) the density at the edge of the gravitational sphere ($r = r_{\rm g}$) at the epoch of $t=0.2$\,Myr  as $\rho_{\rm \infty}=\rho_{\rm rg,tend}$. 
We confirmed that the Bondi accretion rate for density (1) is always higher than that for density (2).
Thus, the former would give a lower limit of the Bondi accretion rate ($\dot{M}_{\rm b, min}$) and the latter is an upper limit ($\dot{M}_{\rm b, max}$).
We consider that our procedure can roughly estimate the Bondi accretion rate at each epoch. 

%%%%%%
% Fig. 7
%%%%%%
\begin{figure*}
  \begin{tabular}{cc}
  	\begin{minipage}[t]{0.5\hsize}
     \centering
     \includegraphics[width=8cm]{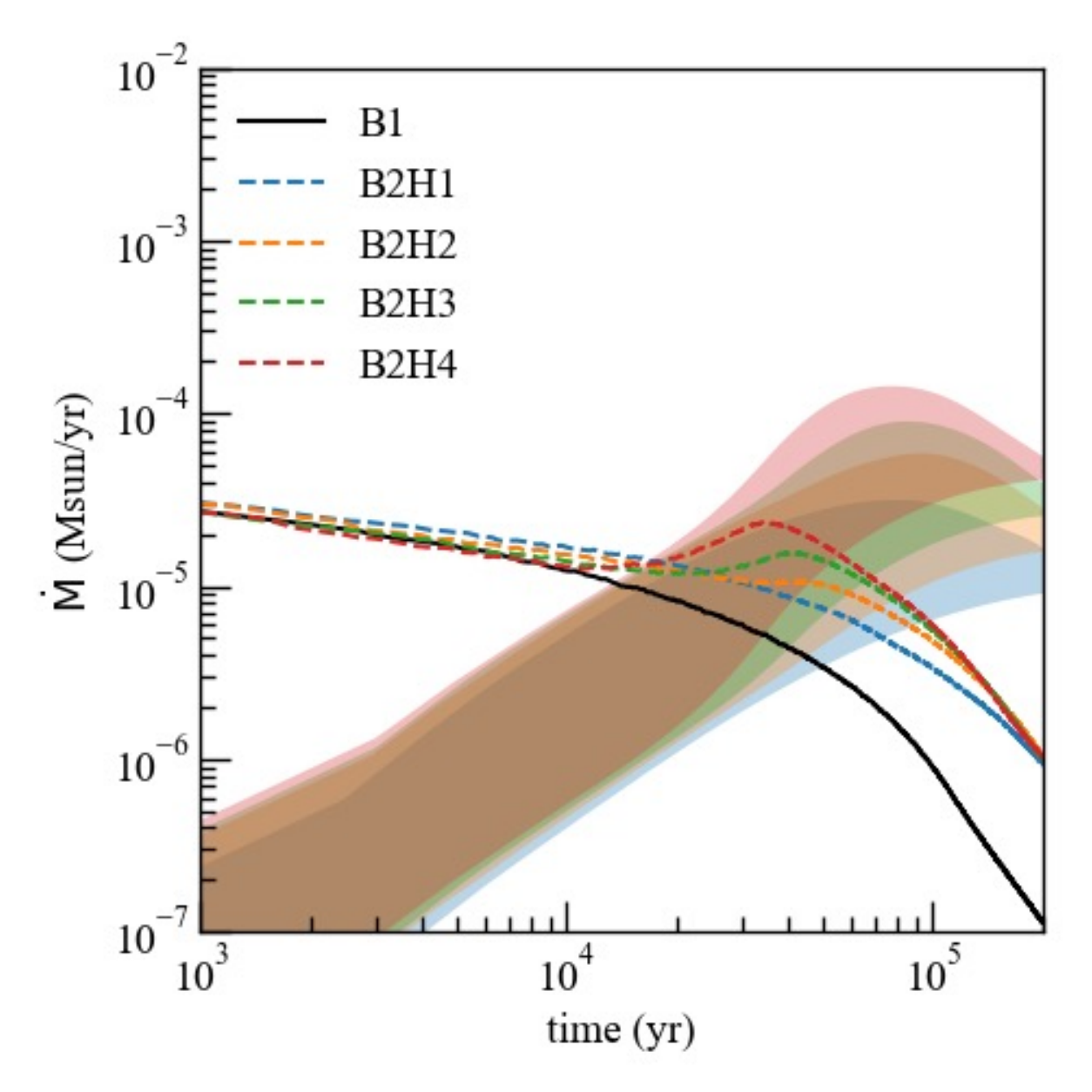}
        \label{tmdt2BEbondirangehigh}
     \end{minipage} &
     \begin{minipage}[t]{0.5\hsize}
     \centering
        \includegraphics[width=8cm]{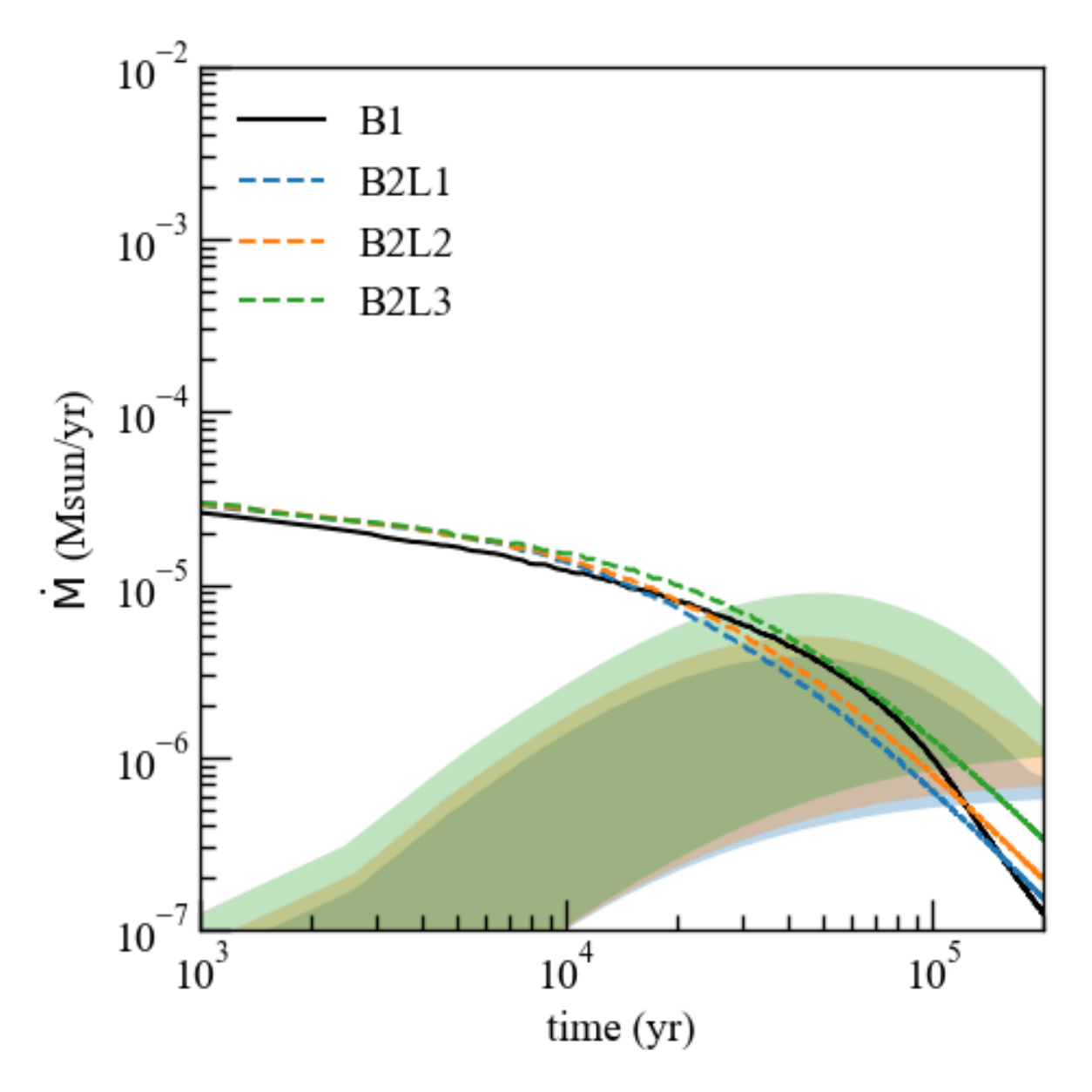}
        \label{tmdt2BEbondirangelow}
     \end{minipage}\\
  	\begin{minipage}[t]{0.5\hsize}
     	\centering
     	\includegraphics[width=8cm]{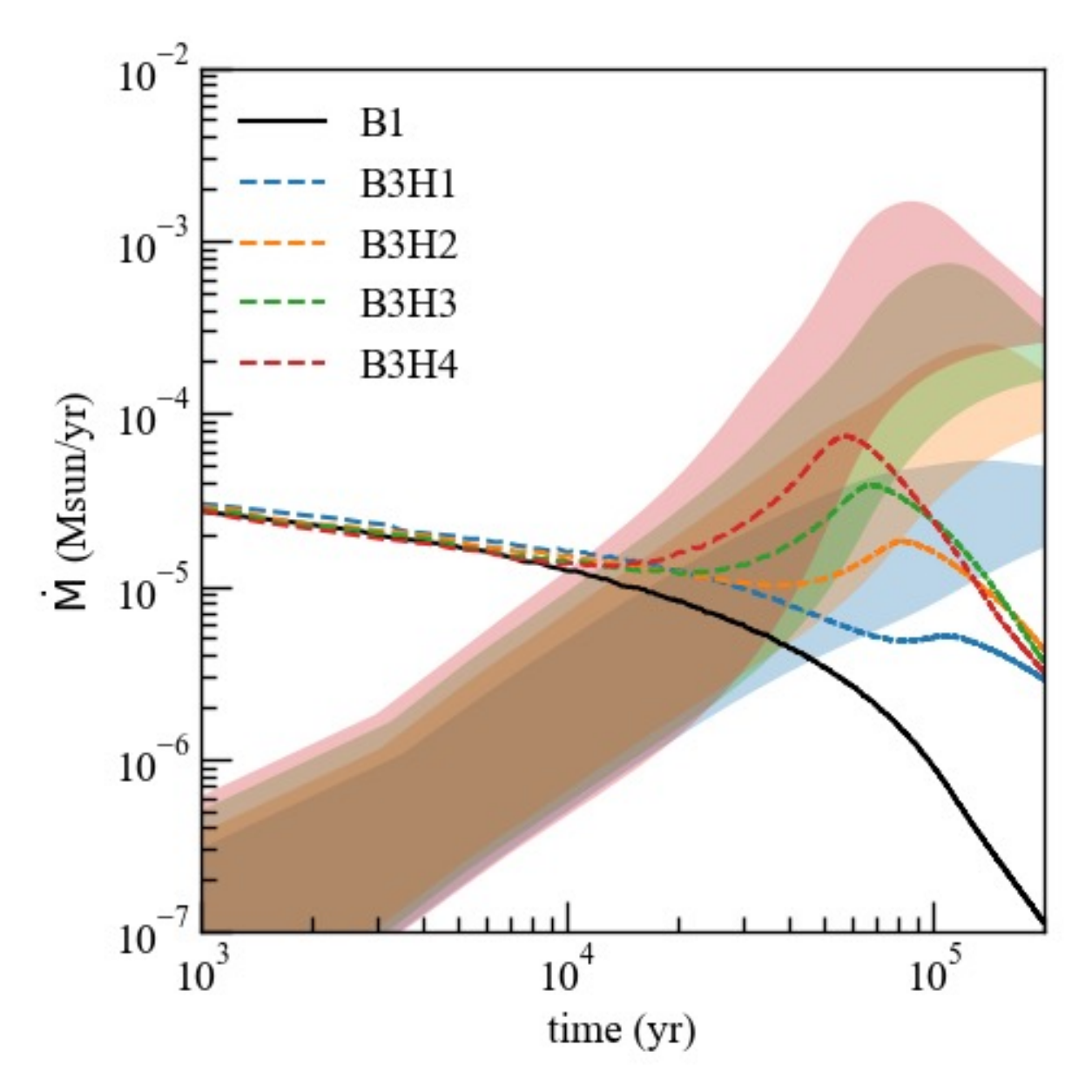}
        \label{tmdt3BEbondirangehigh}
     \end{minipage} &
     \begin{minipage}[t]{0.5\hsize}
     	\centering
        \includegraphics[width=8cm]{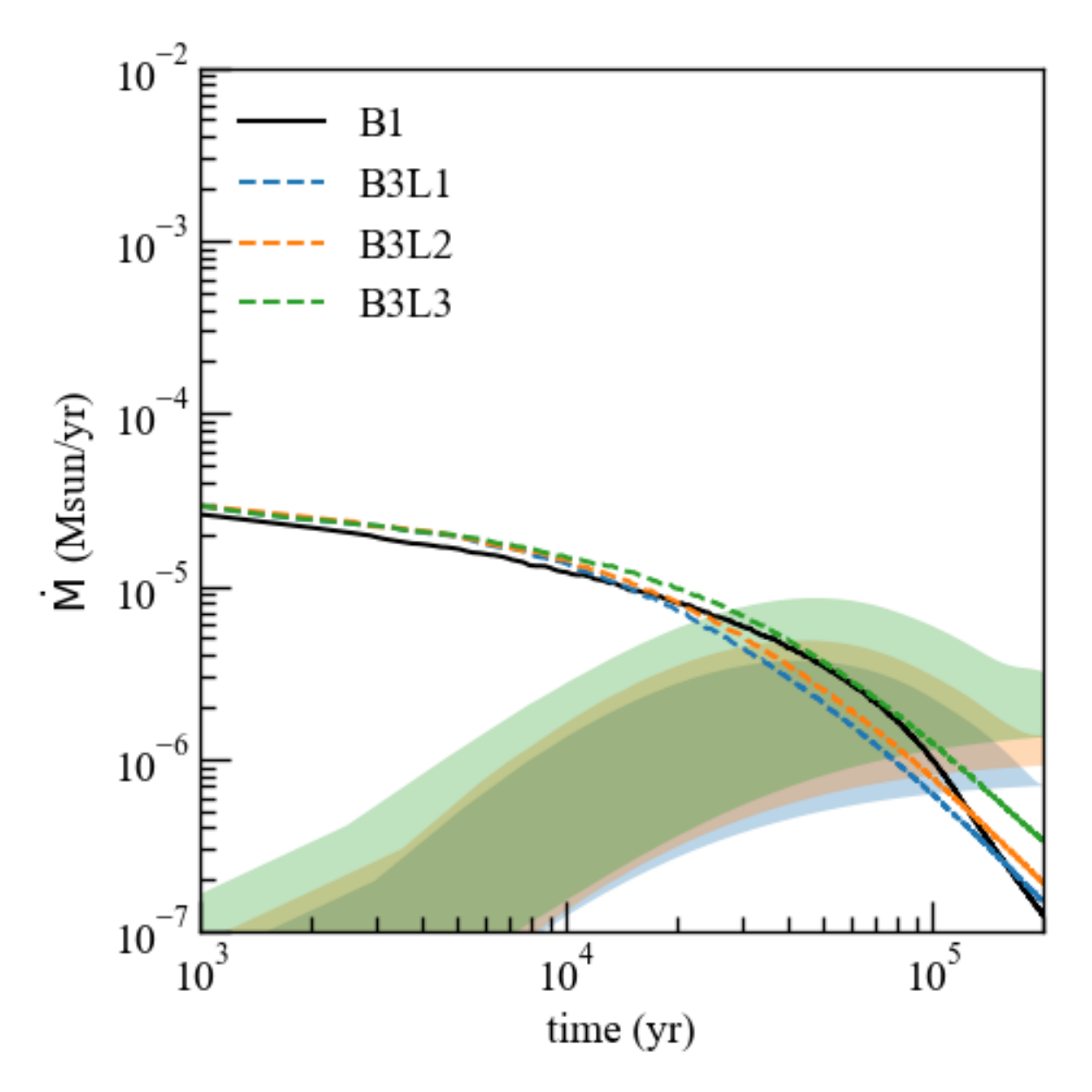}
        \label{tmdt3BEbondirangelow}
      \end{minipage}
    \end{tabular}
    \caption{
Mass accretion rates (broken lines) derived from our simulations plotted over Bondi accretion rates (colored bands) derived from equation~(\ref{bondimdt1}) for all models. 
The colored areas are enclosed by the upper and lower limit of the Bondi accretion rate. 
The broken line and band color are the same for each model. 
The models with a high-density envelope are plotted in the left panels and those with a low-density envelope are plotted in the right panels. 
The models with parameters $r_{\rm g}=2$ and 3 are plotted in the top and bottom panels, respectively. 
The model names are given in each panel. 
Model B1 (black solid line) is plotted in all panels for comparison.
}
\label{tmdtbondirange}
\end{figure*}

Figure~\ref{tmdtbondirange} shows the mass accretion rate for all models except for model B1, along with the Bondi accretion rate given by equation~(\ref{bondimdt1}). 
In the figure, we colored the area  between the lower ($\rho_{\rm \infty}=\rho_{\rm min,t}$ in eq.~[\ref{bondimdt1}]) and upper ($\rho_{\rm \infty}=\rho_{\rm rg,tend}$ in eq.~[\ref{bondimdt1}]) limits of the Bondi accretion rate. 

The Bondi accretion rate for the models with a low density envelope (Fig.~\ref{tmdtbondirange} right panels) continues to increase until $t\sim 5\times 10^4$\,yr. 
The upper limit of the Bondi accretion rate has a peak around  $t\sim 5\times 10^4$\,yr, and then decreases until the end of the simulation. 
On the other hand, the mass accretion rate derived from the simulations continues to decrease, and the protostellar mass does not significantly increase in the later stage. 
In addition, the surrounding density ($\rho_{\rm \infty}=\rho_{\rm min,t}$) continues to decrease (Figs.~\ref{nestedcalc} and~\ref{rvrrho}).
Therefore, the Bondi accretion rate decreases after the protostellar mass growth becomes less significant  (see eq.~(\ref{bondimdt1})).
The lower limit of the Bondi accretion rate is roughly comparable to the accretion rate of the simulations for $t\gtrsim 10^5$\,yr.  
Therefore, for the models with a low-density envelope, the Bondi accretion rate is much lower than the mass accretion rate for the simulations in the early main accretion stage. 
In the later stage, the difference between the Bondi accretion and the mass accretion rates in the simulations is not significant. 
Thus, we do not need to seriously consider Bondi accretion for these models. 

On the other hand, for the models with a high-density envelope, the enhancement of the mass accretion rate in the later main accretion stage can be explained also by Bondi accretion.  
The enhancement of the mass accretion rate is clearer for the models with $r_g=3$ (Fig.~\ref{tmdtbondirange} lower left panel) than for the models with $r_g=2$ (Fig.~\ref{tmdtbondirange} upper left panel). 
As shown in the left panels of Figure~\ref{tmdtbondirange}, the Bondi accretion rate continues to increase and has a peak at $t \sim10^5$\,yr. 
The maximum Bondi accretion rate exceeds the accretion rate in the simulations at $t\sim2\times10^4$\,yr,  
for which both the maximum and minimum Bondi accretion rates are much larger than the accretion rate in the simulations.
The Bondi accretion rate for model B3H1 is about one order of magnitude higher than that for model B2H1. 
In addition, the difference in the Bondi accretion rates among the models with $r_{\rm g}=3$ and a high-density envelope (Fig.~\ref{tmdtbondirange} left bottom panel) is significant. 
For these models, the mass accretion rate is high for a long duration (Fig.~\ref{tmdtbondirange}) and the protostellar mass continues to increase (Fig.~\ref{tm}).  

The difference in the mass accretion rate among the models should be attributed to the surrounding density, which determines the Bondi accretion rate (see eq.~(\ref{bondimdt1})) and resultant protostellar mass. 
Thus, for the models with a high density envelope, it is natural that the mass accretion rate in the later phase  traces the Bondi accretion rate, which is proportional to the square of the protostellar mass (eq.~(\ref{bondimdt1})) when the protostar is massive enough. 
The enhancement of the mass accretion rate means that the protostar can acquire mass from the envelope outside the core within a limited time when the surrounding density is sufficiently high.  
Both the mass accretion rate in the simulations and the Bondi accretion rate begin to decrease in the final stage of the simulations.
The descrease of the Bondi accretion and its applicability is explained in \S\ref{sec:applicability}.

\subsection{Cores within a filament}
\label{sec:cores}
As described in \S\ref{sec:scale}, we can scale our simulation results by changing the central density in the initial cloud core. 
However, recent observations \citep{tokuda2020} imply that gravitational collapse begins when the central density of the prestellar cloud core reaches $n_c\simeq 10^6\,\cc$.  
Thus, we adopted $n_c\simeq 10^6\,\cc$ to construct the critical B.E. sphere.  
The radius of the critical B.E. sphere is $\sim0.02$\,pc with $n_c\simeq 10^6\,\cc$. 
In addition, we set gravitational radii of twice ($r_{\rm g}=2$) and three times ($r_{\rm g}=3$) the critical B.E. radius, in which $r_{\rm g}$ is normalized by the critical B.E. radius (see, \S\ref{sec:initial}).
The dimensional radius of the star-forming core for the models with $r_{\rm g}=2$ is $\sim0.04$\,pc, and that for the models with $r_{\rm g}=3$ is $\sim0.08$\,pc. 
Thus, the diameter of the cores is $d= 0.08$\,pc ($r_{\rm g}=2$) and $0.16$\,pc ($r_{\rm g}=3$), respectively. 
 
Observations of various star-forming regions imply a universal filament width of $\sim 0.1$\,pc \citep{arzoumanian2011, andre2014, arzoumanian2019, tokuda2019,andre2022}. 
%%Note that the universality of the filament width has not been clearly explained in theoretical studies and thus is still controversial even in observational studies. 
The diameters of the star-forming cores adopted in this study ($d= 0.08$\,pc for the models with $r_{\rm g}=2$ and  $d=0.16$\,pc for the models with $r_{\rm g}=3$) are comparable to a filament width of $\sim0.1$\,pc.
The density of filaments is high in a molecular cloud or star-forming region, and the star-forming cores are embedded in the filament. 
Thus, a high-density envelope is expected to enclose a (star-forming) core. 

In addition, theoretical studies \citep[e.g.][]{inutsuka1992} have predicted that fragmentation occurs in filaments with separations of four times the diameter of the cylinder or filament width, indicating that the neighboring cores are separated by about four times the core radius.
Thus, from a theoretical point of view, a gravitational radius comparable to a few times the core radius can be justified \citep[for observations, see, e.g.][]{andre2014,kim2022}.

A filament may have internal structures such as fibers and an anisotropic density distribution \citep[e.g.][]{mizuno1995,hacar2013}.
Thus, it is not simple to model the environment around cores embedded in a filament. 
However,  it would be valuable to understand the environmental effects around the core in the star formation process with simple settings and untangle the relation between the CMF and the IMF. 

\subsection{Effects of protostellar outflow}
\label{sec:outflow}
This study ignored the feedback effects of protostellar.
%% outflow because we adopted unmagnetized non-rotating cores as the initial state. 
Although the outflow strength, such as momentum flux and kinetic energy, depends somewhat on the magnetic field strength and rotation degree of the initial core  \citep{machida2013}, the outflow can reduce the star formation efficiency to $30$--$70$\,\% \citep{machida2012}.
As described above, we imposed neither a magnetic field nor rotation for the initial cloud core, to better focus on the effects of the environment on the mass accretion rate.
With this setting, we could show that a protostar can acquire its mass from outside the cloud core (or from the outer envelope) within a limited time when the  density of the outer envelope is high.

However, in this setup, almost all the accreting matter is converted into the protostar without feedback. 
In reality, outflow expels a significant fraction of the accreting matter from the center to interstellar space \citep{matzner1999,matzner2000}. 
In a future study, by comparing the mass accretion rate derived in this study with that including the outflow feedback, we will  address the effect of outflow on the mass accretion process and protostellar mass growth.

%%It is considered that the outflow mass ejection rate is proportional to the mass infall (or accretion) rate \citep{matsushita2017,matsushita2018}.
%%Previous studies have shown that the outflow gradually weakens when the mass accretion rate continues to decrease with time \citep{machida2013}. 
%%However, the mass accretion rate is temporarily enhanced when the density of the outer envelope is high. 
%%In addition, the histories of the mass accretion rate differ among the models with different outer envelopes. 
%%Thus, it is difficult to predict the mass ejection rate due to the outflow because the outflow behavior would be different for different mass accretion rates.
%%In a future study, we will calculate the evolution of magnetized and rotating cloud cores for different environmental densities to clarify the star formation efficiencies and the relation between the IMF and the CMF.

\subsection{Difference between this and previous studies and interpretation of recent observations}
The enhanced mass accretion rate shown in our study cannot be confirmed based on previous studies. 
The difference between our study and previous studies is the density profile in the outer region of the prestellar core.
The density profile for the outer region that is usually adopted is proportional to $\propto r^{-2}$ (i.e., B.E. sphere) or $\propto r^0$ (i.e., a uniform sphere) because previous studies considered gravitational collapse of an isolated star-forming core.   
In contrast, this study focuses on the environmental effects in the region around star-forming cores on the mass accretion rate. 
Although the calculation methods are almost the same as in previous studies, we assume different circumstances for isolated star formation motivated by recent observations \citep{takemura2021b,takemura2021a},
whereas previous studies did not focus on the envelope around an isolated cloud core. 
It is difficult to investigate the environmental effects of each star-forming core  in analytical studies \citep[][]{myers2011} or global star formation simulations \citep{haugbolle2018,kuffmeier2019,pelkonen2021} because of the highly complex nature of the star-forming environment. 

One of our aims is to connect the studies on isolated star formation to star cluster formation studies. 
We also investigate the environmental effects on the mass accretion rate. 
Although the calculation settings in our study are similar to previous studies, the circumstances we consider differ from previous studies and are more realistic.

\cite{takemura2021b,takemura2021a} showed that the CMF is the same as the IMF and observed a star formation efficiency of about 100\%. 
Since half of the core mass is ejected by outflow, these observations indicate that a mass comparable to the core mass is supplied from the outer envelope and contributes to the growth of protostars. 
Our studies show that when the core has a high-density envelope, a mass comparable to the prestellar core flows into the protostellar system. 
Thus, both the observations and our results indicate that the prestellar cores in the Orion Molecular Cloud region are embedded in a dense envelope. 

\section{Summary}
We investigated the mass accretion rate onto protostars formed in star-forming cores embedded in various environments. 
The star-forming cores were composed of two parts: an inner core with a critical B.E. radius and an outer envelope with a nearly uniform density. 
We limited the region where the gas can accrete to within two or three times the critical B.E. radius to mimic realistic conditions for localized star formation.
We changed the ambient gas density and gravitational radius to investigate the effect of the environment on the mass accretion rate.
We constructed 15 different star-forming cores with different surrounding densities ($n_{\rm ext}=(0.11$--$110) \times 10^4\,\cc$) and gravitational radii ($0.0217$--$0.0652$\,pc). 
We imposed neither rotation nor a magnetic field on the star-forming cores to focus only on the environmental effects. 
Thus, we ignored feedback due to protostellar outflow in this study, unlike in our previous studies \citep[e.g.][]{machida2013, matsushita2017}. 

We calculated the evolution of these star-forming cores  until $2\times10^5$\,yr after protostar formation.
%% using a nested grid code with a sink method.  
The mass accretion rates derived from our simulations agree well with those derived from self-similar solutions in the very early stage.  
The mass accretion rate gradually decreases when the density of the outer envelope is very low, which is consistent with previous numerical studies investigating the mass accretion rate in isolated cloud cores. 

On the other hand, the mass accretion rate temporarily increases in the intermediate or later accretion stage when a high-density gas envelope encloses the (centrally condensed) core, which is assumed to mimic clustered star-forming regions such as the Orion Nebula Cluster.
We assume that we  identify a star-forming core as a core with a centrally condensed region without considering a uniformly distributed density envelope, as is done in observations. This will, however, underestimate the actual prestellar core mass directly related to the CMF.
In such a case, the gas flows into the protostellar system from both a centrally condensed core and the outer envelope.
Thus, the matter in the core and outer envelope contribute to protostellar mass growth.
Therefore, we could not adequately estimate the star formation efficiency, defined as the mass ratio of the star to that of the prestellar core, because the matter flowing from the outer envelope is ignored. 
The origin or precursor of the protostar could not be explicitly identified in observations. 
It may therefore be  difficult to relate the IMF to the CMF in such a clustered star-forming region because we cannot adequately identify the true prestellar core (or actual prestellar matter). 

We also estimated the mass ratio of the (density-enhanced) prestellar core to the protostellar mass, in which the prestellar core is defined as the region within the critical B.E. radius, which does not include the outer envelope. 
We showed that the mass ratio is in the range $\sim0.9$--$2.7$ for the models with $r_{\rm g}=2$  and $\sim0.9$--$7.5$  for the models with $r_{\rm g}=3$. 
Since we did not consider feedback due to protostellar outflow and permitted gas inflow from the outer envelope within the gravitational sphere, the (apparent) star formation efficiency defined in this study exceeds $90$--$750$\%. 
Even if the outflow ejects 70\% of the accreting mass, the star formation efficiency can be estimated to be $\sim30$--$200$\% for the models with $r_{\rm g}=2$  and $30$--$560$\% for the models with $r_{\rm g}=3$.  
Our results indicate that it is difficult to predict the CMF and real star formation efficiency from observations, especially in clustered star-forming regions. 

The main aim of this study was to explain the recent observation by \cite{takemura2021b,takemura2021a} 
that the CMF has the same distribution as the IMF, meaning that the star formation efficiency is about 100\%.  
However, since part of the accreting gas should be expelled by the outflow, their study implies that the prestellar core mass identified in their observations is insufficient to reproduce the IMF in the Orion Nebula Cluster region.
Thus,  an additional mass inflow outside of the identified prestellar cores is required to form the stars in this region. 
It is difficult to observe the ambient or uniform mass envelope using interferometric observations. 
In observations, it is also difficult to identify the actual prestellar core because we could not observe or monitor the whole accretion phase.  
Thus, the high-density envelope surrounding the core, which contributes to the protostellar mass growth, might be missed in identifying prestellar cores in observations. 

If the protostellar outflow ejects 70\% of the accreting mass \citep{andre2010}, about three times the core mass is required to reproduce the IMF in this region.  
In other words, the core to protostar mass ratio should exceed $\sim300$\,\% when considering the outflow feedback.
In our simulation, the (apparent) star formation efficiency, shown in Figure~\ref{sfe}, roughly reaches or exceeds 300\% for models B2H4, B3H2, B3H3 and B3H4. 
These models have a high-density envelope outside the centrally condensed core. 
Thus, our study indicates that the IMF in this region may be reproduced when the centrally condensed cores, identified as prestellar cores in observations,  are embedded in a high-density medium.

This study showed that the mass accretion rate does not monotonically decrease with time when a centrally condensed core is embedded in a high-density envelope. 
For all models, the mass accretion rate in the early stage can be represented by a self-similar solution and is consistent with the results of previous numerical studies. 
On the other hand, the enhancement of the mass accretion rate in the intermediate or later stage  is roughly represented by the Bondi accretion rate.
However, in any case, the mass accretion rate finally decreases to $\ll 10^{-7}\,\mdot$ and protostellar growth stops within $t\lesssim 2\times10^5$\,yr. 
Our study indicates that it is difficult to identify the actual prestellar core (or prestellar mass) because the gas outside the core can contribute to protostellar mass growth depending on the ambient density and distance between cores. 
However, realistic simulations including outflow feedback are required to accurately relate the CMF to the IMF. 

\section*{Acknowledgement}
We thank the referee for very useful comments and suggestions on this paper. 
We have benefited greatly from discussions with Kazuki Tokuda.
This research used the computational resources of the HPCI system provided by the Cyber Science Center at Tohoku University and the Cybermedia Center at Osaka University (Project ID: hp200004, hp210004,hp220003).
Simulations reported in this paper were also performed by 2020, 2021 and 2022 Koubo Kadai on the Earth Simulator (NEC SX-ACE) at JAMSTEC. 
The present study was supported by JSPS KAKENHI Grant (JP17H06360, JP17K05387, JP17KK0096, JP21H00046, JP21K03617: MNM).

\section*{DATA AVAILABILITY}
The data underlying this article are available on request to the corresponding author.

\bibliographystyle{mnras}
\bibliography{nozaki_machida_2022.bib}

\appendix
%%%%%%%%%%%%%%%%%%%%%%%%%%%%%%%%%%
\section{Limitation and Applicability of Bondi accretion}
\label{sec:applicability}
%%%%%%%%%%%%%%%%%%%%%%%%%%%%%%%%%%
The accretion rates in the simulations plotted in the left panels of Figure~\ref{tmdt} decrease after they reach a peak.  
It is considered that there are two reasons for the decline in the accretion rate. 
One is the depletion of the mass in the infalling envelope. 
Since the envelope density continues to decrease with time, the Bondi accretion rate decreases as long as the protostellar mass does not increase. 
As seen in Figure~\ref{tm}, the protostellar mass does not significantly increase for $t \gtrsim 10^5$\,yr. 
%%Thus, the Bondi accretion rate should decrease after the protostellar mass does not significantly increase.

\begin{figure*}
\begin{tabular}{cc}
  	\begin{minipage}[t]{0.5\hsize}
%%     \centering
	\includegraphics[width=8cm]{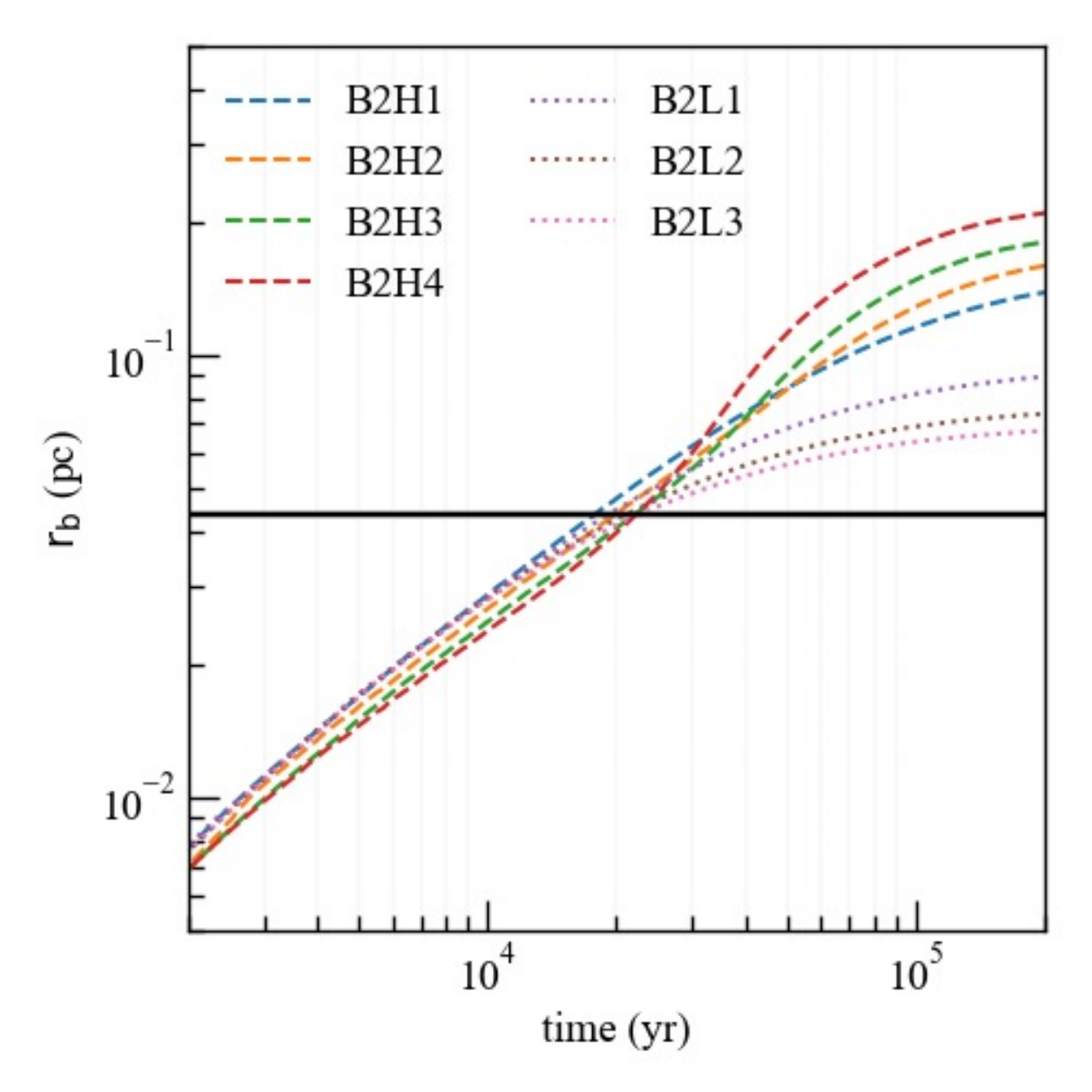}
     \end{minipage}&
     \begin{minipage}[t]{0.5\hsize}
%%	\centering
	\includegraphics[width=8cm]{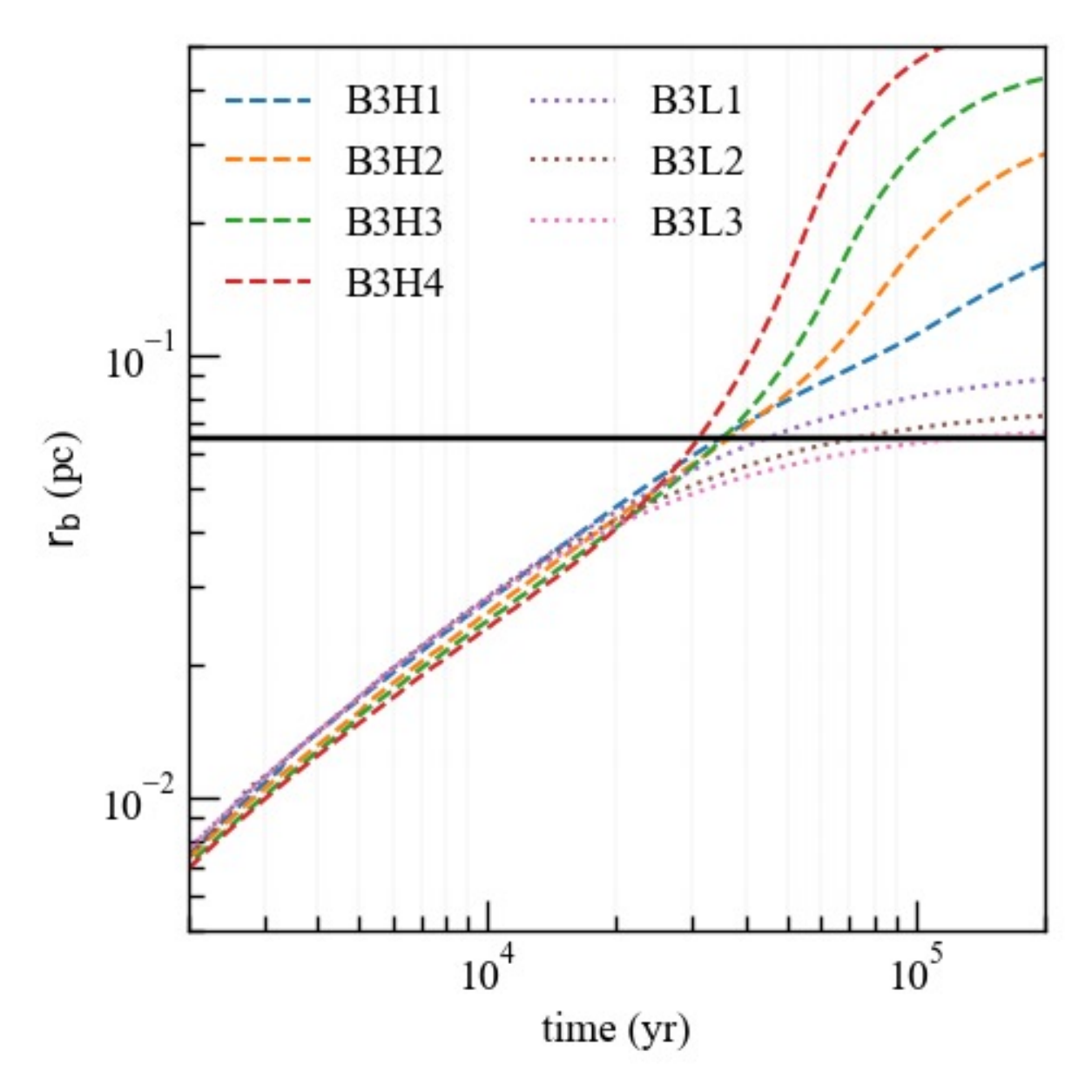}
	\end{minipage}
   \end{tabular}
   	\caption{
Bondi radius  against elapsed time after protostar formation for models with $r_{\rm g}=2$ (left) and 3 (right).  
The black horizontal line in each panel is  the dimensional gravitational radius, $r_{\rm g,pc}=0.0435$\,pc (left) and 0.0652\,pc (right), respectively. 
}
\label{bondir}	
\end{figure*}

The other reason for the decline of the accretion rate is the limited range of the accretion zone (or gravitational sphere) we imposed in this study. 
As described in \S\ref{sec:initial}, we consider a fixed gravitational sphere to mimic a realistic star-forming core. 
The Bondi radius $r_{\rm b}$ is proportional to the protostellar mass as 
\begin{equation}
    r_{\rm{b}}=\frac{GM_{\rm ps}}{c_{\rm s,0}^2}. 
\label{rbondi}
\end{equation}
To compare the Bondi radius (eq.~[\ref{rbondi}]) with the gravitational radius $r_{\rm g}$ (see, \S\ref{sec:initial}), the time variation of the Bondi radius for all models except model B1 is plotted against the elapsed time after protostar formation in Figure~\ref{bondir}. 
The figure shows that the Bondi accretion radius exceeds the gravitational radius at $t\sim2\times10^4$\,yr for the models with $r_{\rm g}=2$ (Fig.~\ref{bondir} left) and  $t\sim3\times10^4$\,yr for the models with $r_{\rm g}=3$ (Fig.~\ref{bondir} right). 
For the models with a low-density envelope, the ratio of the Bondi to the gravitational radius at the end of the simulation is within a factor of 2 for models B2L1, B2L2 and B2L3 (Fig.~\ref{bondir} left panel) and within a factor of 1.3 for models B3L1, B3L2 and B3L3 (Fig.~\ref{bondir} right panel). 
Thus, these models should mostly consist of a contribution due to Bondi accretion (or accretion from the envelope within the Bondi radius). 
On the other hand, the ratio exceeds $2$--$4$ for the models with a high-density envelope (models B2H1, B2H2, B2H3, B2H4, B3H1, B3H2, B3H3, B3H4). 
Thus, for these models, Bondi accretion cannot be fully included in the mass accretion (rate) because only part of the gas within the Bondi radius can accrete onto the protostar. 

Figure~\ref{bondir} also shows that the epoch at which the Bondi radius exceeds the gravitational radius corresponds well with the increase of the mass accretion rate shown in Figure~\ref{tmdt}, especially in the models with a high-density envelope. 
This indicates that Bondi accretion cannot be ignored in these models. 
%%%%%%
% Fig. 8
%%%%%%

\begin{figure*}
\begin{tabular}{cc}
  	\begin{minipage}[t]{0.5\hsize}
     \centering

	\includegraphics[width=8cm]{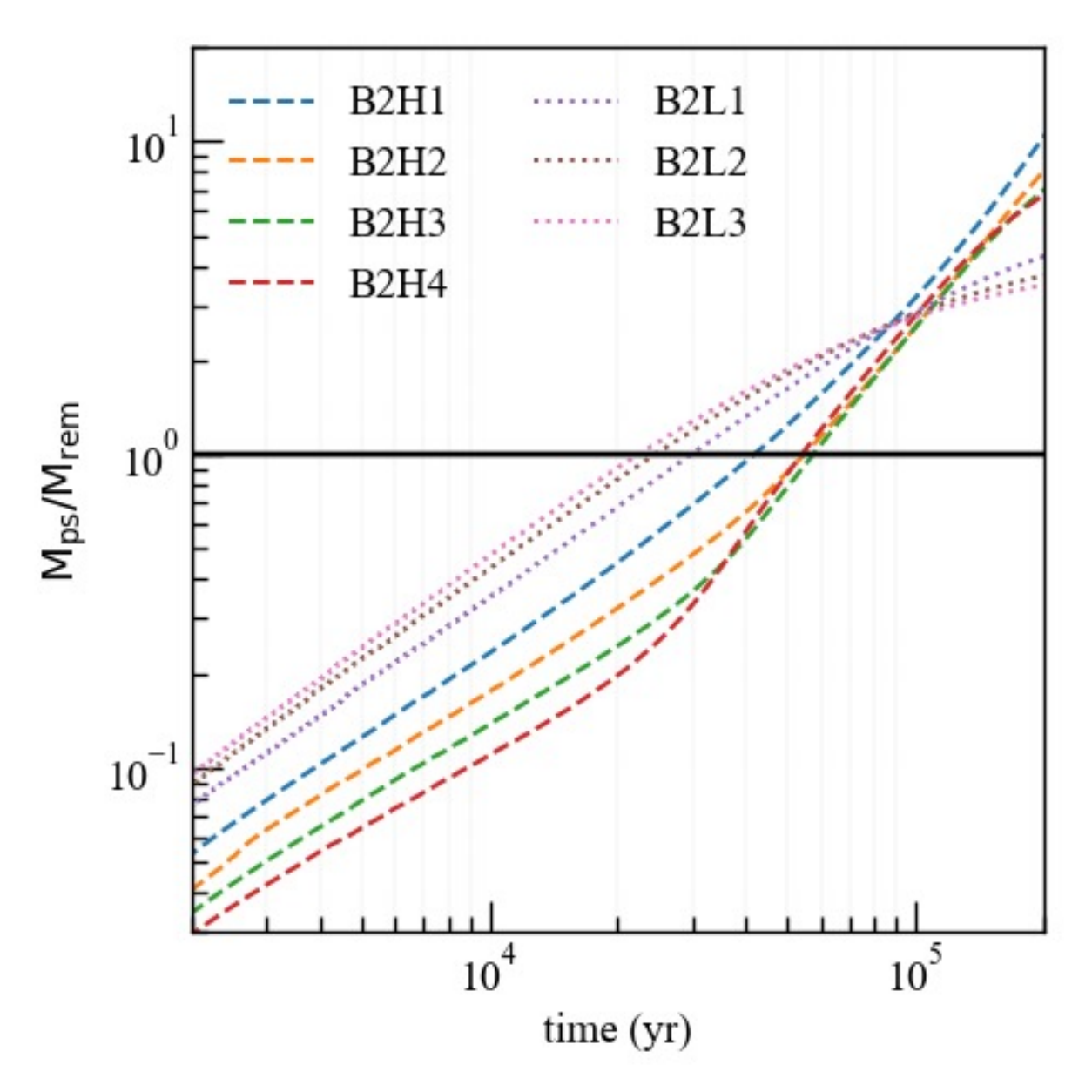}
	\label{tm2BEsink}
     \end{minipage}&
     \begin{minipage}[t]{0.5\hsize}
	\centering
	\includegraphics[width=8cm]{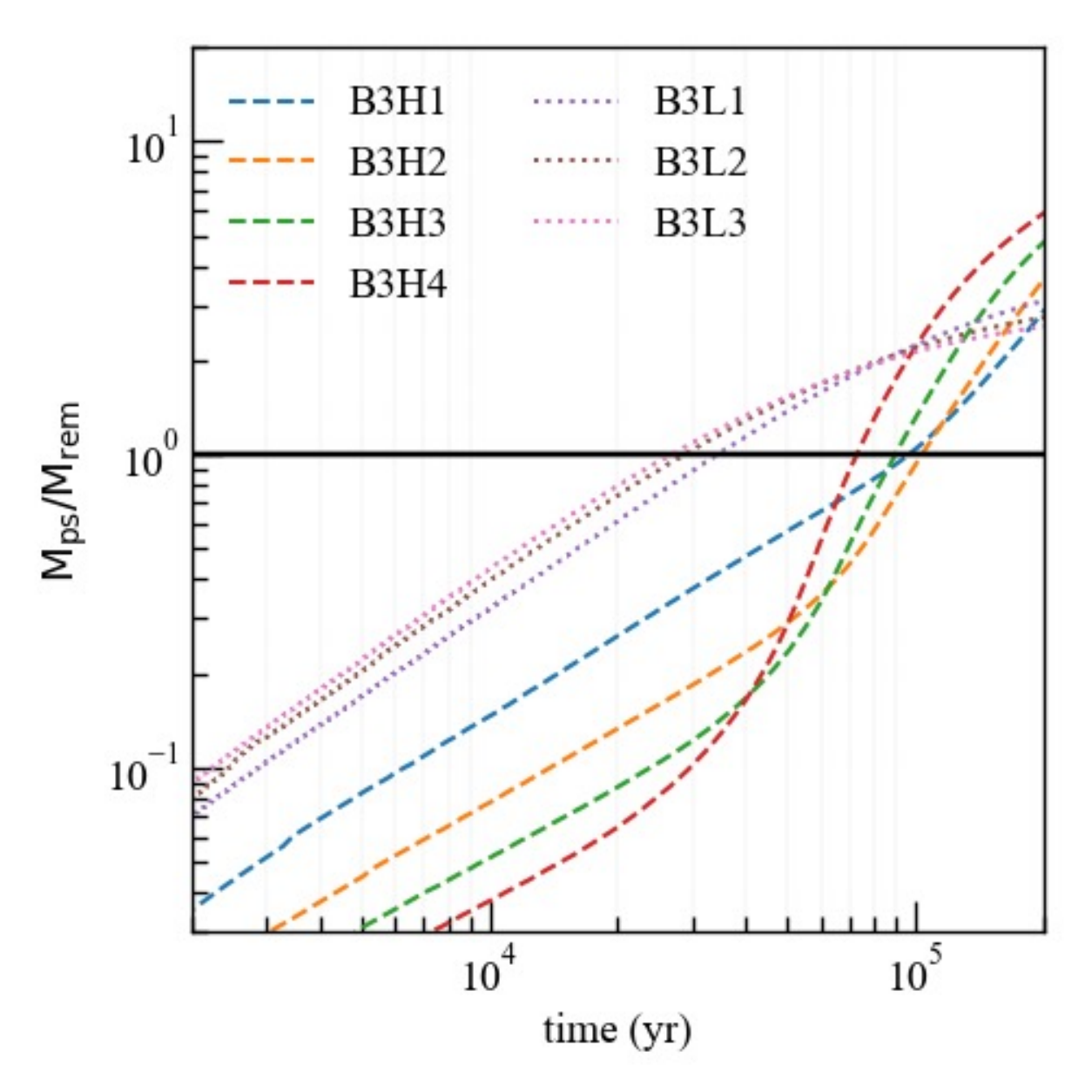}
	\end{minipage}
   \end{tabular}
   	\caption{
Ratio of protostellar mass $M_{\rm ps}$ to remaining mass of star-forming core $M_{\rm env}$ against elapsed time after protostar formation for models with $r_{\rm g}=2$ (left) and 3 (right). 
}
\label{ratio}	
\end{figure*}

Finally, we discuss the applicability of Bondi accretion.
A point mass or point gravity source is assumed when considering Bondi accretion. 
In contrast, both the masses of  a protostar (point mass) and self-gravitating core are considered in deriving self-similar solutions of gravitationally collapsing cloud cores \citep{larson2003}. 
Since the self-similar solutions cannot represent the long-term evolution of star-forming cores, we considered Bondi accretion as an alternative to discuss the mass accretion rate in later stages.  
The self-gravity of the star-forming core is ignored when considering Bondi accretion.
Thus,  it is helpful to compare the protostellar mass $M_{\rm ps}$ with the remaining mass in the star-forming core $M_{\rm rem}$. 
Figure~\ref{ratio} plots the ratio $M_{\rm ps}/M_{\rm rem}$, with $M_{\rm rem}$ calculated as
\begin{equation}
M_{\rm rem} = \int_{r<r_{\rm g}} \rho \, dV. 
\end{equation}
We  note that Bondi accretion should be applicable when $M_{\rm ps} \gg M_{\rm rem}$. 
%%%%%%
% Fig. 9
%%%%%%
Figure~\ref{ratio} shows that the ratio $M_{\rm ps}/M_{\rm rem}$ increases with time because the protostellar mass increases, but the remaining  mass of the star-forming core decreases with time. 
The protostellar mass exceeds the remaining mass ($M_{\rm ps} > M_{\rm env}$) at $t\simeq(4$--$6)\times10^4$\,yr for the models with $r_{\rm g}=2$ and $t=(7$--$10)\times10^4$\,yr for the models with $r_{\rm g}=3$. 
However, Figure~\ref{tmdt} shows that the enhancement of the mass accretion rate begins at $t\sim (2$--$4)\times 10^4$\,yr for the models a high-density envelope. 
Thus, both accretion mechanisms (self-similar solution of collapsing core and  Bondi accretion) could be considered to represent the mass accretion rate found in this study more realistically.

Figure~\ref{ratio} also shows that the protostellar mass greatly dominates the remaining mass for $t\gtrsim10^5$\,yr. 
Thus, the mass accretion rate in the simulations can be represented by the Bond accretion rate at such later stages. 
Figure~\ref{tmdt} shows that the mass accretion rates for the models with a high-density envelope are higher than those for the models with a low-density envelope for $t>10^5$\,yr.   
The difference is likely due to the difference in the density of the outer envelope, which determines the Bondi accretion rate at later stages. 

%%%%%%%%%%%%%%%%%%%%%
\subsection{Scalable physical quantities}
\label{sec:scale}
%%%%%%%%%%%%%%%%%%%%%
We use an isothermal equation of state as described in \S\ref{sec:initial}. 
Thus, we can rescale the physical quantities by changing the initial central density $n_{\rm c}$ of the star-forming core. 
With a central number density of $n_c= 10^6\,\cc$, we can derive physical quantities such as radius and mass, as described in \S\ref{sec:initial}.
We also use the physical quantities estimated with $n_c= 10^6\,\cc$ in the following. 

However, the physical quantities can be rescaled with $n_c$ by rescaling the unit length $l_{\rm unit}$, unit mass $m_{\rm unit}$  and unit time $t_{\rm unit}$ as 
$l_{\rm unit}=l_{n6}/(n_{\rm{c}}/10^{6}\,\cc)^{1/2}$,  $M_{\rm unit}=M_{n6}/(n_{\rm{c}}/10^{6}\,\cc)^{1/2}$ and $t_{\rm unit}=t_{n6}/(n_{\rm{c}}/10^{6}\,\cc)^{1/2}$, where $l_{n6}= 6.96\times10^2$\,au, $M_{n6}=2.21\times10^{-3}\,\msun$ and $t_{n6}=1.75\times10^4$\,yr are the unit length, unit mass and unit time for $n_{\rm c}=10^6\, \cc$. 
The mass accretion rate does not depend on the initial central density because it has the dimensions of [M/T] (mass divided time) and the term $(n_{\rm{c}}/10^{6}\,\cc)^{1/2}$ is canceled.

\end{document}